\documentclass[10pt,journal,cspaper,compsoc]{IEEEtran}

\usepackage{graphicx, color}
\usepackage{subfigure}
\usepackage{framed}
\usepackage{balance}
\usepackage{epsfig}
\usepackage{times}
\usepackage{rotating}
\usepackage{listings}
\usepackage{amssymb}
\usepackage{url}
\usepackage{hyperref}
\usepackage[table]{xcolor}
\usepackage{array}
\usepackage{ifthen}
\usepackage{soul}
\usepackage{multirow}
\usepackage[disable]{todonotes}
\usepackage{float}
\usepackage{tikz}
\usepackage{booktabs}
\usepackage{dirtytalk}
 \DeclareGraphicsExtensions{.eps}

\ifCLASSOPTIONcompsoc
\else
\fi
\ifCLASSINFOpdf
\else
\fi

\begin{document}
\title{Need for Sleep: the Impact of a Night of Sleep Deprivation on Novice Developers' Performance}

\author{
Davide Fucci, Giuseppe Scanniello,~\IEEEmembership{Member,~IEEE}, Simone Romano, and  Natalia Juristo
\IEEEcompsocitemizethanks{
\IEEEcompsocthanksitem D. Fucci is with HITeC and the University of Hamburg, Hamburg, Germany \protect\\ E-mail: fucci@informatik.uni-hamburg.de 
\IEEEcompsocthanksitem G. Scanniello and S. Romano are with DiMIE - University of Basilicata, Potenza,  Italy. \protect\\
E-mail: \{giuseppe.scanniello, simone.romano\}@unibas.it
\IEEEcompsocthanksitem N. Juristo is with  Technical University of Madrid, Madrid, Spain \protect\\
E-mail: natalia@fi.upm.es}
\thanks{}}

\markboth{Submission to Transactions on Software Engineering}
{Fucci \MakeLowercase{\textit{et al.}}: Title}

\IEEEcompsoctitleabstractindextext{
\begin{abstract}
We present a quasi-experiment to investigate whether, and to what extent, sleep deprivation impacts the performance of novice software developers using the agile practice of test-first development (TFD). 
We recruited 45 undergraduates, and asked them to tackle a programming task. 
Among the participants, 23 agreed to stay awake the night before carrying out the task, while 22 slept normally. 
We analyzed the quality (i.e., the functional correctness) of the implementations delivered by the participants in both groups, their  engagement in writing source code (i.e., the amount of activities performed in the IDE while tackling the programming task) and  ability to apply TFD (i.e., the extent to which a participant is able to apply this practice).
By comparing the two groups of participants, we found that a single night of sleep deprivation leads to a reduction of 50\% in the quality of the implementations. 
There is notable evidence that the developers' engagement and their prowess to apply TFD are negatively impacted. 
Our results also show that sleep-deprived developers make more fixes to syntactic mistakes in the source code.
We conclude that sleep deprivation has possibly disruptive effects on software development activities. 
The results open opportunities for improving developers' performance by integrating the study of sleep with other psycho-physiological factors in which the software engineering research community has recently taken an interest in.
\end{abstract}

}

%
\maketitle 

\IEEEdisplaynotcompsoctitleabstractindextext

\IEEEpeerreviewmaketitle

\section{Introduction}\label{Intro}
The loss for the American economy due to the lack of sleep of its workforce, whether chronic or not, is estimated to be 63 billion USD every year.\footnote{\url{http://ti.me/xXa17G}}
Nevertheless, as depicted in a scene from the movie \textit{The Social Network} where Mark Zuckerberg writes code for 36 hours straight, forgoing sleep appears to be a badge of honor in the programmers and start-up communities.

The importance of sleep is nowadays recognized in the field of economics, where it is shown that sleep disturbances contribute to decreased the employees' performances at a high cost for the employers (e.g.,~\cite{rosekind2010cost, vidavcek1986productivity}). In management science, sleep loss was found to bewilder decision-makers activities~\cite{harrison1999one}.

In general, the lack of sleep affects working memory, creativity, decision making, multitasking ability, response time, and focus~\cite{pilcher1996effects}.
Not getting enough sleep prevents the brain from restoring its effectiveness, as it needs to work harder to accomplish the same amount of work~\cite{dinges1995overview}. 
In particular, the performance for tasks that require attention declines as a function of hours of sleep deprivation~\cite{9781429242288}.

In recent years, the software engineering community has been interested in the role played by factors related to human biology and physiology concerning several aspects of software development (e.g.,~\cite{siegmund2014understanding,muller2015stuck,fritz2014using}). 
A survey of 311 developers~\cite{sarkar2017characterizing} found that \textit{sleepiness} is perceived as one of the main causes of mental fatigue resulting in performance drop. 
However, \textit{how} sleep affects software developers has not been studied so far. 
Given the link between sleep and cognitive performance showed in physiological research (e.g.,~\cite{dinges1995overview,linde1992effect}), we believe that sleep deprivation can have serious repercussions for software developers' performance.
The assessment of the effects of sleep deprivation will contribute to the broader investigation of the role of physiological factors for software engineers, with the goal of supporting them in their work; for instance by informing them when to take breaks so to avoid counterproductive actions. 

In this paper, we investigate the following primary research question:
\begin{framed}
\noindent
To what extent does sleep deprivation impact developers' performance?
\end{framed}
To answer this research question, we performed a quasi-experiment with 45 (final-year)  Undergraduate Students in Computer science at the University of Basilicata in  Italy. The participants in the experiment worked on a programming task requiring them to use the popular agile practice of test-first development (TFD)~\cite{beck2003}; 23 of them did so while being totally sleep deprived---i.e., they forewent sleep the night.  
We based our experiment on TFD because, together with unit testing, it was the main focus of the course in which the experiment was embedded. Moreover, TFD is well known and largely applied in software industry, its application requires discipline and rigor~\cite{martin2007professionalism,jeffries2007guest,one20159th}, and there exists non-invasive, validated tools that measure whether the process is followed correctly ~\cite{Becker2015494}.

Participants were assigned to the treatment group (i.e., students who forewent sleep the night before the experiment)  and to the control group (i.e., students who slept regularly the night before the experiment) by their availability to forgo one night of sleep, instead of a random assignment. Therefore, we consider our investigation a quasi-experiment rather than a randomized controlled experiment. 

Comparing the two groups of participants, we found that:
\begin{itemize}
  \item the quality of the software measured as functional-correctness produced by sleep-deprived software developers drop by half compared to developers under normal-sleep condition;  
\item sleep deprivation can have an impact on the engagement of  developers, as well as their ability to follow the TFD practice.
\end{itemize}

The first contribution of this paper is to present and discuss the results of the first empirical study on the role of sleep deprivation in the software engineering field.
Overall, the results of our quasi-experiment suggest that assessing sleep condition can provide indications  on the quality of the source code that software developers write as well as their performance. 
Our second contribution is a replication package \footnote{\url{https://doi.org/10.6084/m9.ﬁgshare.5483974}} and a series of lessons learned to foster replications and further studies. 
\textbf{Paper Structure.}  In Section~\ref{sec:Background}, we  provide background information and present related work. 
In Section~\ref{sec:Design}, we show the design of our experiment.  The findings are reported in  Section~\ref{sec:Findings}, and discussed in Section~\ref{sec:Discussion}.
Final remarks and future work conclude the paper in  Section~\ref{sec:FinalRemarks}. 


\section{Background and Related Work}\label{sec:Background}
In this section, we report an overview of the software engineering research involved with the study of the biological and physiological facets of software developers (Section~\ref{sec:psico}) and the key concepts used in our study from the medical research about sleep (Section~\ref{sec:sleepWork}).

\subsection{Physiological factors software engineering}\label{sec:psico} 
Investigating human cognitive endeavor through physiological measures is nowadays a standard practice in psychology~\cite{andreassi2013psychophysiology, kramer1991physiological}.
In recent years, the software engineering research community has taken an interest in studying how several aspects of software development (e.g., code comprehension~\cite{siegmund2014understanding}, 
software quality~\cite{muller2016using}) impact the software developers' cognitive state---measured using physiological signals.
Some of the first studies in this context are in the sub-field of program comprehension.
For example, Sharif and Maletic~\cite{sharif2010eye} used an eye-tracking device to understand developers' naming conventions strategies. 
In a within-subjects controlled experiment, the authors observed differences in visual effort and elapsed time between developers comprehending source code identifiers written using camel case and the ones who read code written using underscore (i.e., snake case). 
They recommend novices to use the latter style, although the gap between the two narrows with experience. 
Similarly, Bednarik and Tukiainen~\cite{bednarik2006eye} used eye-tracking to understand what part of source code expert developers look at, compared to novices. 

Other investigations aims to understand the developers' brain.
Floyd \textit{et al.}~\cite{floyd2017decoding} used medical imagining, through functional Magnetic Resonance Imaging (fMRI), to assess which areas of the brain are activated when reading source code.
They found that the same areas of the brain are activated when reading source code as well as when reading prose text. 
However, different cerebral activities are associated with the two types of text.
Moreover, their experiment showed that more experienced developers tend to read prose and code similarly at the neural level. 
Siegmund~\textit{et al.}~\cite{siegmund2014understanding} explored the feasibility to use fMRI to directly measure program comprehension.
To that end, the authors conducted a controlled experiment with 17 participants, with the same level of programming experience, recruited among computer science and mathematics students. 
Participants were asked to comprehend six short source code snippets. Siegmund~\textit{et al.} found distinct activation patterns of five brain regions related to working memory, attention, and language processing~\cite{siegmund2014understanding}. 
The results of a different controlled experiment using fMRI on program comprehension were reported in~\cite{Siegmund:FSE:2017}. 
The authors involved 11 participants and manipulated experimental conditions to isolate specific cognitive processes related to bottom-up comprehension and comprehension based on beacons providing hints about the purpose of a snippet (e.g., method signatures and common programming idioms). 
The results showed that beacons ease comprehension. 
Ikutani and Uwano~\cite{ikutani2014brain} measured developers' brain activity when comprehending code.
They used Near Infra-Red Spectroscopy (NIRS), a less invasive alternative to fMRI which measures the brain frontal-pole activity.
Their small scale experiment shows that such area is activated when comprehending source code in which a variable needs to be memorized, but not when parsing \textit{if-else} statements. 
Parnin~\cite{parnin2011subvocalization} used subvocal utterances, emitted by developers while performing two tasks of different complexities. 
The author found a statistically significant difference between the number and intensity of subvocalization events corresponding to the two tasks.
Fritz \textit{et al}.~\cite{fritz2014using} combined three physiological features (i.e., eye movement, the electrical signals of skin and brain) to distinguish programming tasks according to their difficulty.
These results showed the potential of physiological techniques for characterizing development tasks according to their difficulty.

Developers' performance recently became the subject of research exploiting physiological measures, too.
For example, Radevski~\cite{radevski2015real} proposed a continuous monitoring of developers productivity based on brain electrical activities; whereas M\"uller and Fritz~\cite{muller2015stuck} used an ensemble of physiological metrics to measure the progress and incorruptibility of developers performing small development tasks. 

M\"uller and Fritz~\cite{muller2016using} investigated the use of biometrics to identify code quality concerns in real time. In a longitudinal experiment with graduate students---then replicated on a smaller scale with professional software developers---they showed that heart rate, respiration rate, and skin temperature could determine the parts of the codebase in which developers are more likely to introduce a bug. Compared with this investigation, our quasi-experiment can be considered complimentary because we study the effect of sleep deprivation---usually associated with reduced blood flow in several regions of the brain, and changes in body temperature~\cite{thomas2000neural}---on the capacity of developers to write source code.

Software development, like many other intellectual activities, is ruled by emotions that can result in stressful situations (e.g., pressing deadlines, work within a restricted budget)~\cite{Wastell:1993tq}. Ostber \textit{et al.}~\cite{Anonymous:oFOdivcs} proposed the use of salutogenesis\footnote{It focuses on factors that support human health and well-being, rather than on factors that cause disease~\cite{Antonovsky1979}. Specifically, this model is concerned with the relationship between health, stress, and coping.} to support stressed developers. 
In their experiment, the authors make use of the concentration of cortisol and $\alpha$-amylase in the saliva, as the physiological measure to infer the participants' stress level.

Sarkar and Parnin analyzed mental fatigue of software developers~\cite{sarkar2017characterizing}. 
They surveyed 311 software developers and carried out an observational study with nine professionals.
Their results show that fatigued developers have problems in focusing, coming up with optimal solutions and tend to make logical mistakes causing bugs. 
Moreover, fatigue hampered developers' creativity and motivation. 
They reported that one of the leading cause of mental fatigue for software developers is \textit{sleepiness}.

Our research differs from previous work because our goal is to directly collect evidence about the role of sleep deprivation in software development. This study is the first in this respect. 

\subsection{Physiology of sleep} \label{sec:sleepWork}
The daily hours of sleep needed vary depending on factors such as age and gender, with the average sleep duration being between seven and eight and half hours per night~\cite{weaver2007relationship}.

The empirical, medical research on sleep deprivation focuses on the effects of forcing the participants to sleep less than usual by keeping them awake between 24 and 72 hours~\cite{alhola2007sleep}.
These kinds of studies are very laborious and expensive to carry out leading to compromises in the study design~\cite{alhola2007sleep}. 
For example, a small sample size can reduce the statistical power of the experiment, but a larger population may come at the expense of other methodological issues, such as a reduction in the cognitive test selection~\cite{alhola2007sleep}. 

Medical research has shown that sleep deprivation decreases cognitive performance because of the wake-state instability due to microsleeps---i.e., very short periods of sleep-like state~\cite{durmer2005neurocognitive}.
Moreover, sleep deprivation negatively affects the reactivity to stimuli from emotions~\cite{pilcher2015sleep}. 
The non-medical scientific communities studied the effects of sleep deprivation mainly in the field of manufacturing (e.g.,~\cite{dinges1995overview,vidavcek1986productivity}) and decision making (e.g.,~\cite{harrison1999one,harrison2000impact}), but not in the software engineering field. 
Sleep deprivation has detrimental effects on central features necessary for software development~\cite{bergersen2011programming}, such as working memory---that part of short-term memory in charge of manipulating transient information, which is fundamental for problem solving~\cite{linde1992effect,alhola2007sleep}---attention, and decision making.
Table~\ref{tbl:cognitiveSD} reports the main effects of sleep deprivation on  cognitive performances that can have consequences for software development.



\begin{table}[t]
\centering
\caption{Summary of cognitive performance effects of sleep deprivation}
\label{tbl:cognitiveSD}
\begin{tabular}{ll}
\toprule
\textbf{Cognitive state induced by sleep deprivation}&\textbf{References}\\
\midrule
Adverse mood changes &~\cite{taub1973performance,totterdell1994associations,david1991rapid}\\
\midrule
Loss of situational awareness &~\cite{monk2007practical}\\ 
\midrule
Reduced learning acquisition &~\cite{yoo2007deficit,rasch2013sleep}\\ 
\midrule
Deterioration of divergent thinking &~\cite{horne1988sleep,goel2009neurocognitive}\\ 
\midrule
Perseveration on ineffective solutions &~\cite{durmer2005neurocognitive,monk2007practical}\\ 
\midrule
Behavior compensatory effort &~\cite{engle2014effects,hockey1998effects}\\ 
\bottomrule
\end{tabular}
\end{table}

There are several approaches to empirically measure sleep deprivation. 
These include self-assessment (e.g., on a scale with values from ``completely sleepy'' to ``completely awake''~\cite{kaida2006validation}), sensor-based (e.g., smart band, polysomnography~\cite{douglas1992clinical}, PET scan), and psychometrics such as the psychomotor vigilance task (PVT)~\cite{dinges1985microcomputer,drummond2005neural}.
PVT, employed by NASA to monitor astronauts' sleep condition~\cite{dinges2012psychomotor}, is cheaper and easier to administer compared to sensor-based approaches, while its measurements converge to the ones of more sophisticated tests~\cite{dinges1985microcomputer}. 
A PVT task lasts for ten minutes during which the subject needs to react (e.g., by pressing a button on a keyboard) to a visual stimulus (e.g., a symbol appearing on a blank screen) before a given time threshold (usually 500 millisecond)~\cite{Basner2011}.
An error (i.e., failing to react to the stimulus) or long reaction time (i.e., the timeframe between the stimulus appearing on the screen and the button being pressed) is attributable to attention lapses due to micro-sleep events indicating a condition of sleep deprivation~\cite{Basner2011}. 
The number of errors and the reaction time (RT) are then used to compute the following metrics~\cite{Basner2011}:
\begin{itemize}
  \item \textit{Performance score}: The percentage of correct reactions to stimuli; 
  \item \textit{Mean 1/RT}: The mean of the reciprocals of all RT;
  \item \textit{10\% Slowest 1/RT}: The mean of the reciprocals of the slowest 10\% RT;
  \item \textit{Minor Lapses}: The number of errors;
  \item \textit{10\% Fastest RT}: The mean of the fastest 10\% RT.
\end{itemize}
The reciprocal transform (1/RT) is one of the PVT outcomes most sensitive to total and partial sleep loss~\cite{Basner:2011:Sleep}. This metric emphasizes slowing in the optimum and intermediate response and it substantially decreases the contribution of long lapses, which is why the slowest and the fastest 10\% of RTs are usually reciprocally transformed.

In this study, we use PVT, on top of a self-assessment questionnaire, to assess the adherence to the treatment---i.e., whether a participant slept or not. In the medical field, PVT is  used not only to measure sleep deprivation~\cite{dinges1985microcomputer,drummond2005neural,doi:10.1177/1541931215591181} but also to perform data cleaning~\cite{10945,44348}.
We followed these good practices and used both PVT and the participants' self-declaration to get a more accurate assessment. 

The human-computer interaction research community has previously used PVT to support  the  development of context-aware  applications.  
For example, Abdullah \textit{et  al.}~\cite{AMM16} conducted  an  ecological momentary assessment about the mobile phone usage of 20 participants over 40 days. 
The participants were prompted to take several PVT tasks during the day on their smartphone. 
By correlating the mobile phone usage with the PVT scores, the authors were able to predict, in an unobtrusive fashion, the  participants’ cognitive  performance  and  alertness at  a given  time  of  the  day.  With  such  information,  a  software system can improve its users productivity---e.g., by suggesting them to tackle the most important tasks while at the peak of their alertness. 
Leveraging a mechanism similar to PVT (i.e., measuring a subject’s  physical  reaction  to  a  visual  stimuli),  Althoff \textit{et  al.}~\cite{AHW17}  performed  the  largest  study  on  the  impact  of sleep  (and  lack  of  thereof)  on cognitive  performance. 
The  authors triangulated the sleep quality measurement, obtained from wearable devices, of more than 31.000 US-based Microsoft product users with their interactions with the Bing search engine\footnote{https://www.bing.com} (75 millions keystrokes and mouse click). 
Their results align with laboratory-based sleep studies where PVT is  used.  
They  show  that  a  single  night  of  partial  sleep deprivation (i.e., less than six hours) increases reaction time up to 4\% with respect to normal sleeping condition (i.e., seven to nine hours).  
They also show that a  decrease  in  cognitive  performance due  to  two  consecutive  nights  of  partial  sleep  deprivation can  last  up  to  a  period  of  six  days  before  it  is  fully recovered.



\begin{table}[t]
\centering
\caption{Summary of the experimental settings}
\resizebox{0.5\textwidth}{!}{
\begin{tabular}{ll}
\toprule
\multicolumn{1}{c}{Variable} & \multicolumn{1}{c}{Value}                      \\ \midrule
Participants                     & 45 Computer Science Undergraduate Students \\
Groups size                   &  22 regular sleep/23 sleep deprivation (15 after the removal based on PVT)            \\
(control/treatment)  & \\
Development env.      & Java 8, JUnit 4, Eclipse 4.4.2 \\
Training                     & Information System (IS) course \\
& (12 hours in lab, home assignments)               \\
Experimental task            & PigLatin (8 confirmations)                     \\
Task duration                & 90 minutes                                     \\
Date                         & 9 a.m. December 12, 2015                       \\
Place                        & DiMIE, University of Basilicata (Italy)                \\ \bottomrule
\end{tabular}
}

\label{tbl:summary}
\end{table}

\section{Study Design}\label{sec:Design}
We performed a quasi-experiment following the recommendations provided by  Juristo and Moreno~\cite{juristo}, Kitchenham \textit{et al.}~\cite{kitchenham02preliminary}, and Wohlin \textit{et al.}~\cite{wohlin12}. In Table~\ref{tbl:summary}, we  show a summary of the experimental setting of this quasi-experiment ---e.g., main variables representing the context of the study.

In the following, we first define  the goal of our quasi-experiment and present the research questions (Section~\ref{sec:Goal}). Successively, we present the independent and dependent variables of our study (Section~\ref{sec:variable}) and show the hypotheses defined to investigate the research questions  (Section~\ref{sec:hypotheses}). We provide details on the participants in the experiment (Section~\ref{sec:context}), present its design (Section~\ref{sec:design}) and the infrastructure used to collect and analyze data (Section~\ref{sec:AnalysisProc}). We conclude by highlighting the experiment operation (Section~\ref{sec:operation}) and experimental object (Section~\ref{sec:tasks}). 


\subsection{Goal}\label{sec:Goal}
Following our main research question presented in Section~\ref{Intro}, we defined the main goal of our quasi-experiment by applying the Goal Question Metrics (GQM) template~\cite{basili94} as follows:

\vspace{+0.3cm}
\noindent
\textbf{Analyze} developers' sleep deprivation  \\
\textbf{for the purpose of} evaluating its effects \\
\textbf{with respect to} the quality of produced source code, the developers' engagement with the task, and their ability to follow TFD \\
\textbf{from the point of} view of the researcher\\
\textbf{in the context of} an Information System (IS) course involving novice developers and students in Computer Science and Software Engineering.\\
\vspace{+0.3cm}

\begin{figure*}[ht]
\centering
  \includegraphics[width=.93\textwidth]{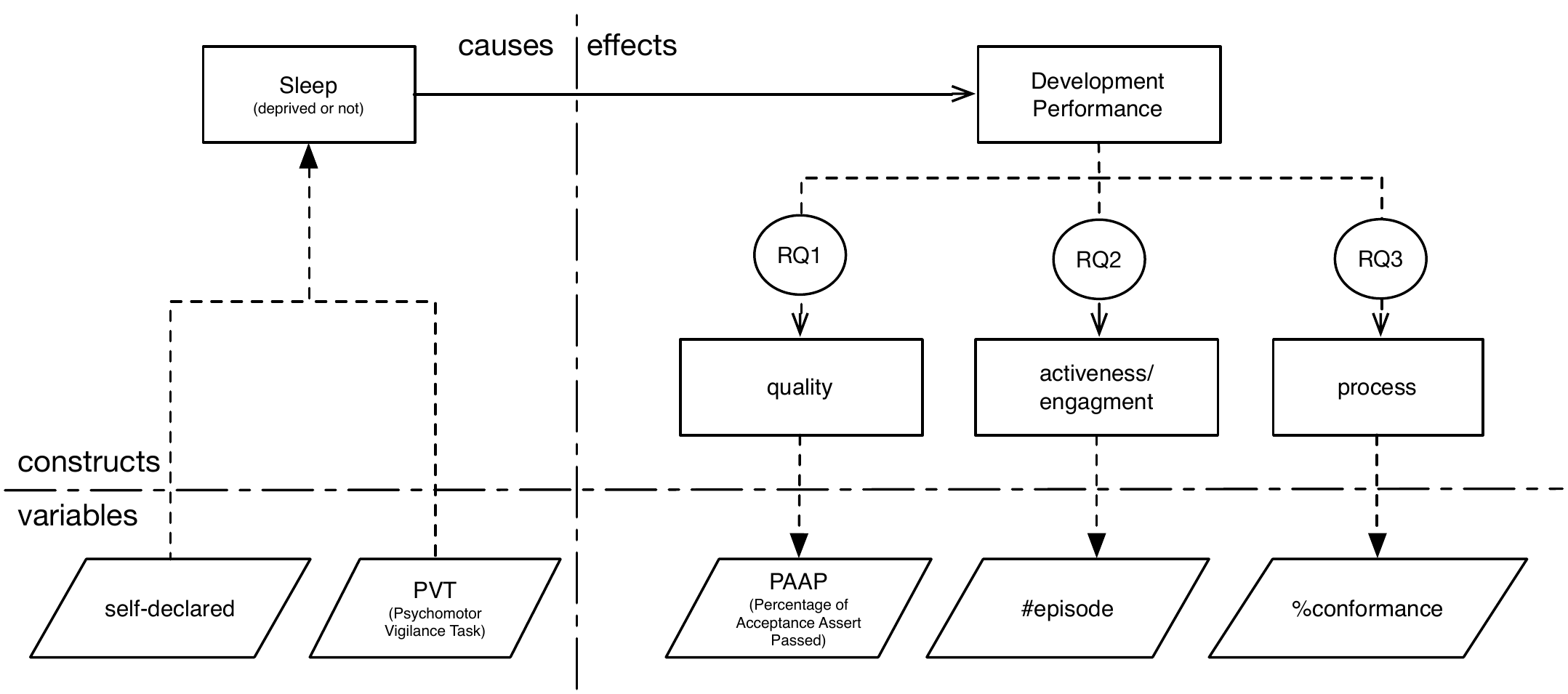}
  \caption{Conceptual model and operationalization of our quasi-experiment}
\label{fig:ConceptualModel}
\end{figure*}

Accordingly, we defined and investigated the following research questions:
\begin{description}
\item[RQ1.] Does sleep deprivation decrease the quality of the solution to a programming task?
\item[RQ2.] Does sleep deprivation decrease the developers' engagement with a programming task? 
\item[RQ3.] Does sleep deprivation decrease the ability of developers to apply TFD to a programming task?
\end{description}
The conceptual model and the operationalization of the constructs investigated in this experiment are presented in Figure~\ref{fig:ConceptualModel}. The rectangles in the upper part of this figure show the experiment constructs and their relationships with the research questions (in the ellipses). The bottom part shows the instruments used to allocate the treatments (left-hand side) and the metrics to measure the constructs (right-hand side).

\subsection{Variables selection}\label{sec:variable}
Given the experiment design, the independent variable is \textit{sleep}, a nominal variable with two levels, regular sleep ($RS$) and sleep deprived ($SD$). 
The control group ($RS$) includes the participants who slept normally the night before carrying out the experimental task, while the treatment group ($SD$) includes the participants who forewent sleep the night before the experimental task.

The three dependent variables deal with three different constructs: quality, engagement, and process. 
Quality is intended here as a measure of how well the software satisfies the functional requirements.
The metric we used is the percentage of acceptance asserts passed ($PAAP$)---the portion of the task correctly implemented based on an acceptance test suite, representing the compliance of the solution to the high-level requirements expressed in the user story. 
The acceptance test suite was developed by the researchers and hidden from the participants. 

In particular, we calculated this metric as follows:
\begin{equation}\label{eq:prod}
\small
PAAP = \frac{\#ASSERT(PASS)}{\#ASSERT(ALL)} \times 100
\end{equation}
$PAAP$ measures the percentage of assert statements in the acceptance test suite passed by the production code delivered by a participant within the fixed duration of the task (i.e., 90 minutes).
From a practical perspective, $PAAP$ represents how well the implementation provided by a participant fits the functional requirements expressed in the acceptance tests. 
The metric is equivalent to the concept of \textit{functional correctness} reported in the ISO/IEC 25010.
The total number of asserts included in the test suite---i.e., $\#ASSERT(ALL)$---is 13.
$PAAP \in [0, 100]$; the higher its value, the better the  quality of the implementation. The definition of this metric is founded on the recommendations to evaluate performance presented in Bergersen \textit{et al.}~\cite{Simula.simula.2934}. 

The construct \textit{engagement} (or also \textit{activeness}) refers to how active the participants are in completing the task.
A proxy to measure such construct is the number of activities a participant performs in the IDE while working.  
The activities we consider are a stream of low-level actions~\cite{kou2010operational} taking place while programming.
The first column of Table~\ref{tbl:types} reports the activities we consider and the related sequence of actions.

\begin{table*}[!ht]
\centering
\caption{Heuristics used by Besouro to infer the development activities (from~\cite{kou2010operational}). Number of recognized activities reported in parentheses.}
\begin{tabular}{ll}
\toprule
Activity                           & Actions sequence                                                                                                                                            \\ \midrule
\multirow{4}{*}{Test-first development (3006)}    & Test creation $\rightarrow$ Test compilation error $\rightarrow$ Code editing $\rightarrow$Test failure $\rightarrow$ Code editing $\rightarrow$ Test pass \\
                               & Test creation $\rightarrow$ Test compilation error $\rightarrow$ Code editing $\rightarrow$ Test pass                                                       \\
                               & Test creation $\rightarrow$ Code editing $\rightarrow$ Test failure $\rightarrow$ Code editing $\rightarrow$ Test pass                                      \\
                               & Test creation $\rightarrow$ Code editing $\rightarrow$ Test pass                                                                                             \\ \midrule
\multirow{3}{*}{Refactoring (772)}   & Test editing (file size changes $\pm$ 100 bytes) $\rightarrow$ Test pass                                                                                                                           \\
                               & Code editing (number of methods, or statements decrease) $\rightarrow$ Test pass                                                                                                                           \\
                               & Test editing AND Code editing $\rightarrow$ Test pass                                                                                                      \\ \midrule
\multirow{2}{*}{Test addition (215)} & Test creation $\rightarrow$ Test pass                                                                                                                          \\
                               & Test creation $\rightarrow$ Test failure $\rightarrow$ Test editing $\rightarrow$ Test pass                                                                 \\ \midrule

\multirow{2}{*}{Production code (16)}& Code editing (number of methods unchanged, statements increase) $\rightarrow$ Test pass \\
								& Code editing (number of methods increase, statements increase) $\rightarrow$ Test pass \\
								& Code editing (size increases) $\rightarrow$ Test pass  \\ \midrule
\multirow{2}{*}{Test-last development (88)}     & Code editing $\rightarrow$ Test creation $\rightarrow$ Test editing $\rightarrow$ Test pass                                                                  \\
                               & Code editing $\rightarrow$ Test creation $\rightarrow$ Test editing $\rightarrow$ Test failure $\rightarrow$ Code editing $\rightarrow$ Test pass                                                                                                                     \\ \bottomrule
\end{tabular}
\label{tbl:types}
\end{table*}

The metric $\#episodes$, which represents the total number of  development activities performed within the duration of the experiment, is used to measure the participants' engagement. 
We considered the sensible development activities (see Table~\ref{tbl:types}) for the specific process and task at hand (i.e., TFD applied to a simple task).    
$\#episodes$ assumes values in the interval $[0, +\infty]$ the larger the number of development episodes (i.e., the higher the value for $\#episodes$), the greater the engagement of a developer.



The \textit{process} construct represents the extent to which a participant is able to follow TFD---i.e., develops a failing unit test, and then implements the production code to make it pass.\footnote{Although refactoring is one of the steps in TFD\@; here we do not explicitly address it due to the simplicity of the task.} 
We study this construct because we have postulated that a night of sleep deprivation could negatively affect how developers adhere to a development technique (i.e., TFD) that requires them to be more focused than the traditional approach to software development. 
The process construct is measured through the $\%conformance$ as follows:
\begin{equation}\label{eq:conf}
  \small
  \%conformance = \frac{\#activities(\textnormal{Test-first development})}{\#activities} \times 100,
\end{equation}
where $\#activities$ is the total number of episodes recognized, while implementing the entire programming task.
In other words, $\%conformance$ measures the percentage of development activities that were recognized as \textit{Test-first development}.
Therefore, $\%conformance \in [0, 100]$; the higher the value, the higher the adherence to TFD. 
We opted for this metric because it is well known and widely used in empirical studies on test-driven development (TDD) and TFD~\cite{fucci2015towards,Hofer:2009kd}.

\subsection{Hypotheses formulation}\label{sec:hypotheses}






Given the literature regarding the effects of sleep deprivation on cognition (see  Section~\ref{sec:Background}), it is expected to have detrimental effects on software developers' performance.  Therefore, we formulated the following null hypothesis to check the effect of sleep deprivation on the quality construct:
\begin{itemize}
\item $H0_{QLTY}$: The quality of source code produced by developers who stayed awake the night before is not better than the quality of source code produced by developers who slept normally. 
\end{itemize}

We formulated the following null hypothesis to check the effect of sleep deprivation on the activeness construct:
\begin{itemize}
\item$H0_{ACTV}$: The activeness during an implementation task of developers who stayed awake the night before is not better than the activeness during the same task of developers who slept normally.
\end{itemize}

Finally, we formulated the following null hypothesis to check the effect of sleep deprivation on the process construct:
\begin{itemize}
\item $H0_{PROC}$: The adherence to TFD during an implementation task by developers who stayed awake the night before is not better than the adherence to TFD during the same task by developers who slept normally.
\end{itemize}

\begin{table*}[!t]
\centering
\caption{Relevant experience levels of the participants (n = 45) in the dataset. Sleep-deprived participants (n=23) reported in parentheses}
\label{tbl:experience}
\begin{tabular}{lccccc}
\toprule
                        & \multicolumn{4}{c}{\textbf{Experience levels}}                                                                                             & \multicolumn{1}{l}{}                 \\ \cmidrule{2-6} 
                        & \multicolumn{1}{l}{Very Inexperienced} & \multicolumn{1}{l}{Inexperienced} & \multicolumn{1}{l}{Neither} & \multicolumn{1}{l}{Experienced} & \multicolumn{1}{l}{Very Experienced} \\ \midrule
Programming (general)   & 0 (0)                                      &  2 (1)                             & 36 (20)                     & 5 (2)                           & 2 (0)                                \\ \midrule
Object oriented programming            & 0 (0)                                     & 6 (3)                             & 33 (18)                     & 5 (2)                           & 1 (0)                                \\ \midrule
Unit testing            & 0  (0)                                    & 38 (20)                           & 6 (3)                       & 1 (0)                           & 0 (0)                                   \\ \cmidrule{2-6} 
Test-first development & 16 (8)                                 & 20 (12)                           & 7 (2)                       & 2 (1)                           & 0 (0)                                   \\ \cmidrule{2-6} 
Eclipse IDE             & 35 (20)                                & 8 (3)                               & 2 (0)                       & 0 (0)                              & 0 (0)                                   \\ \bottomrule
\end{tabular}
\end{table*}

The alternative hypotheses are formulated as follows:
\begin{itemize}
\item$H1_{QLTY}$: The quality of the  source code produced by developers who stay awake the night before is worse than the the quality of source code produced by developers who slept normally (i.e., $QLTY_{SD} < QLTY_{RS}$).

\item$H1_{ACTV}$: Developers who stay awake the night before the execution of an implementation task are less active than developers who slept normally (i.e., $ACTV_{SD} <  ACTV_{RS}$).

\item$H1_{PROC}$: Developers who stay awake the night before the execution of an implementation task adhere less to TFD than developers who slept normally (i.e., $PROC_{SD} < PROC_{RS}$).
\end{itemize}

We have formulated $H0_{QLTY}$, $H0_{ACTV}$, and $H0_{PROC}$ to study RQ1, RQ2, and RQ3, respectively.

\subsection{Sampling and participants}\label{sec:context}

The participants in the experiment were final-year undergraduate students enrolled in an Information System (IS) course in Computer Science at the University of Basilicata (Italy).
The content of the course includes elements regarding software testing, software development processes, software maintenance, and agile development practices with a focus on TFD, regression testing, and refactoring. 
Participants had passed all the exams related to the following courses: Procedural Programming, Object-Oriented Programming I, and Databases.
In these courses, the participants gained experience with C/C++, Java, and TFD.
%
The experiment was conducted as an optional exercise at the end of the IS course. 
We informed the participants that their grade in the IS course would not be affected by their participation in the experiment.
The research questions were not disclosed to the participants until the completion of the experiment. 
The participants were aware that their data would be treated anonymously and disclosed only in aggregated form. 
 
Out of the 95 students enrolled in the IS course, 45 decided to take part in the experiment. 
The results of a pre-questionnaire---administered before the experiment---showed that they had an average experience of 0.5 years as professional programmers. 
In general, the average years of experience with programming (e.g., in university courses, own projects) was 3.6 years; whereas, their experience with software testing was a little over one and a half year. 
The pre-questionnaire was composed of Likert items to be rated on a five-points scale (i.e., from \textit{Very Inexperienced} to \textit{Very Experienced}). The answers are summarized in Table~\ref{tbl:experience}.
The participants  were in good health conditions and aged between 20 and 34 years (average 23.56 years).

\subsection{Experiment design}
\label{sec:design}
The quasi-experiment was designed to have one factor (i.e., \textit{sleep}) and two treatments---sleep-deprived condition ($SD$) and regular sleep condition ($RS$), where the latter is control group~\cite{wohlin12}. 
 22 participants (six females) slept normally  before the experiment, while 23 participants (two females) stayed awake.

We do not consider our study as a controlled experiment because randomization was not possible. 
In experimental design theory, randomization involves the random allocation of experimental units (i.e., participants) to the experimental groups~\cite{shadish2002experimental,wohlin12}.
That is, in a design that uses randomization, the participants have the same chance to be assigned to the control or treatment group.
Randomization can help to reduce the systematic differences between the groups, except for the manipulated variable of interest~\cite{shadish2002experimental}. In such settings, it is possible to identify a strong causal link between the manipulation of the treatments and observed outcomes. 
In our case,  randomization was discouraged due  to  ethical  reason.   
In particular, law and university regulations prevent the possibility to pay students to take part in the study, as well as forcing them to forgo sleep. 
Consequently, the participants are assigned to the experimental groups (i.e., treatment and control) based on their voluntary choice to forgo (or not) sleep for one night before the study took place. 
Although there is some medical evidence that males and females are affected by sleep deprivation in different ways~(e.g.,~\cite{alhola2007sleep,blatter2006gender,ferrara2015gender}), we could not include gender as a blocking factor.
Forcing a balanced number of female and male participants in the experimental groups was not possible because participation was voluntary. 
Similar considerations can be done for the participants’ age as blocking factor.

The pre-existing conditions of the participants are of paramount importance for experiments with human participants~\cite{TuckerDrob:2011ft}.
In our case, we had to address the possibility that participants possessed different pre-existing skills and experience regarding software development. 
To that end, we sampled from a homogeneous population---i.e., students attending the same course, with a similar academic background. 
We assessed the two experimental groups according to their GPA\footnote{In Italy, the exam grades are expressed as integers and assume values between 18 and 30. The lowest grade is 18, while the highest is 30.} (Grade Point Average), used as a proxy to measure the ability of computer science students with software engineering tasks (e.g.,~\cite{TSE2012sd,tseRiccaPTTC10,ScannielloGRTD15}).
The average GPA of the participants in the $RS$ group was 23.92, while for the $SD$ it was 24.3. 
We further analyze the participant's pre-existing skills and experience to substantiate participants homogeneity before the study (see Appendix~\ref{app:1}). 
Accordingly, we assume the differences between groups to be negligible, and participants to be homogeneous. 

We discarded alternative designs, such as repeated measure\footnote{In a repeated measure design, a participant receives one night of sleep or not in two different periods}, because they are more vulnerable to threats to validity prominent in our context.
In particular, a learning effect~\cite{wohlin04} can interfere with the result once the participants' performance are measured under one experimental condition (i.e., regular sleep) and later under the other (i.e., sleep deprivation)
In such settings, the latter measurement can be the result of prior practice under the former condition.
Controlling or compensating for such effect is particularly risky as it implies the use of statistical techniques that hampers the interpretation of the results~\cite{vegas2016crossover,madeyski2017effect}.

Another discarded alternative involved creating of a baseline (i.e., a programming task different from the one used in the experiment) for the SD group \textit{before} the experiment, and then comparing the participants' performances obtained in the two tasks. 
The introduction of possible bias is barely moved from the difference between participant's skills to the difference between tasks.

\begin{figure*}[t]
\centering
\includegraphics[width=1\textwidth]{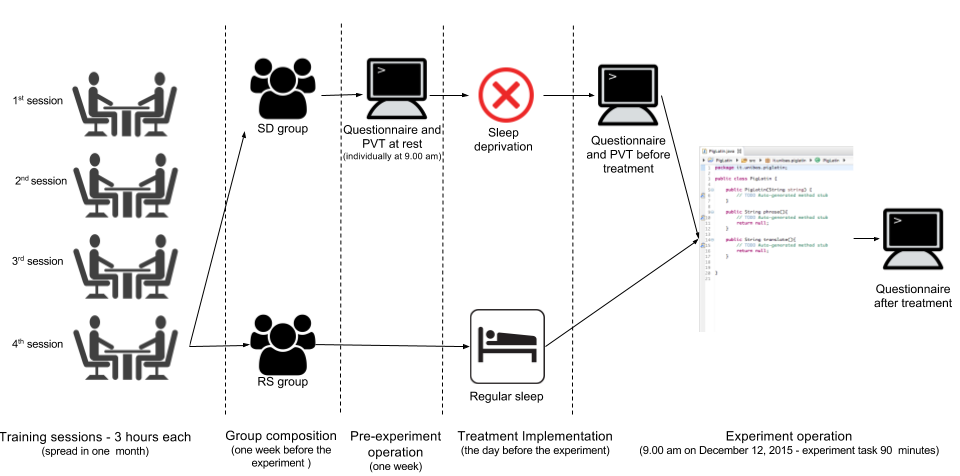}
\caption{Workflow of the study.}§ 
\label{fig:overview}
\end{figure*}

\subsection{Data collection and analysis} \label{sec:AnalysisProc}\label{sec:cleaning}\
To compute $\#episodes$ and $\%conformance$, we leverage Besouro~\cite{Becker2015494}, an Eclipse IDE plug-in able to identify development activities, which are assigned to the following categories: \textit{test-first development}, 
\textit{refactoring}, \textit{test addition}, \textit{production code}, and \textit{test-last development}. 
The identification of these activities is based on the heuristics by Kou \textit{et al.}~\cite{kou2010operational} reported in Table~\ref{tbl:types}. 
The set of heuristics matches sequences of actions (see second column of Table~\ref{tbl:types}) with the information logged by the IDE during its usage.
In total, the plugin registered 4097 activities and 6348 actions. 

As for the adherence of the participants to the treatment, in addition to the method based on the participants self-declaration, (see left hand side of Figure~\ref{fig:ConceptualModel}), a few days before the quasi-experiment, each participant belonging to the $SD$ group filled a questionnaire concerning their perceived quality of sleep during the previous night (pre-experiment operation in Figure~\ref{fig:overview}). 
The participants declaring good quality of sleep took PVT at 9.00 a.m. in a research laboratory at the University of Basilicata. In case a participant perceived that she did not sleep well, we asked her to return at least one day later to retake both the questionnaire and PVT\@. 
This procedure allowed us to obtain baseline PVT scores (i.e., PVT scores at rest). PVT was administered to the participants using a software installed on a PC and a regular keyboard to register the participants' interaction~\cite{khitrov2014pc}.
The day of the experiment, before carrying out the experimental task, each participant in the $SD$ group performed PVT again and filled the same questionnaire concerning their perceived quality of sleep during the previous night (see Figure~\ref{fig:overview}). 

All participants in the $SD$ group carried out the experimental task independently from their PVT scores. 
We used the PVT scores to decide whether a participant in the $SD$ group forewent a night of sleep.  
That is, we exploit PVT scores to assess the adherence to the sleep-deprivation treatment (see  Figure~\ref{fig:ConceptualModel}).
If a participant is sleep-deprived, we expect that her PVT values are worse on the day of the study than at rest.
In Table~\ref{tab:differencesPVT}, we report the differences between the PVT scores at rest and the ones obtained on the day of the study.
Given how the PVT scores are calculated, 
a positive value of the differences for the following metrics  \textit{Performance score}, \textit{Mean 1/RT}, \textit{10\% Slowest 1/RT}  indicates a condition of sleep deprivation. Conversely, a negative value for \textit{Minor Lapses} and \textit{10\% Fastest RT} indicates a condition of sleep deprivation.
We followed a conservative approach to estimate if a participant  stayed awake the night before the experiment. In particular, a participant in the $SD$ group did not stay awake if at least one PVT score  measured the day of the experiment was better than the score at rest. 
Following this criteria, we marked eight participants as non-compliant to the treatment (i.e., the gray rows in Table~\ref{tab:differencesPVT}). 
For example, P7 had one more lapse (i.e., an error) when taking PVT at rest, vis-\'a-vis after the treatment; P7 also gave faster answers after the treatment compared to normal rest condition, as indicated by the 10\% fastest RT values in Table~\ref{tab:differencesPVT}.
Such unexpected PVT score values led us to separately consider P7 and other seven participants during further analysis. 

All the participants in the $SD$ group (included those highlighted in Table~\ref{tab:differencesPVT}) declared that the night before the experiment they were awake between 16 and 24 hours. On average, they did not sleep for 20.7 hours ($\pm$ 4.3 hours) before carrying out the experimental task. 
Participants in the $RS$ group declared an average sleep time of 6.5 hours ($\pm$ 1 hour) the night before that experiment.


To improve the reliability of treatment implementation and reduce possible threats to conclusion validity, we considered two instances for the $SD$ group, namely  \textit{$SD_{Uncleaned}$} and \textit{$SD_{Cleaned}$}. The participants who declared to forego sleep the night before the task were included in the \textit{$SD_{Uncleaned}$} dataset. On the other hand, \textit{$SD_{Cleaned}$} comprised the participants in the $SD$ group after removing the eight participants marked in Table~\ref{tab:differencesPVT}. 
As a result of data cleaning, $SD_{Uncleaned}$ contained all the 23 participants in $SD$, while $SD_{Cleaned}$ included 15 participants. 
The average \textsc{GPA} of the participants in $SD_{Cleaned}$ was 24.37 (vs. 24.3 in $SD_{Uncleaned}$)---the removal of participants did not affect the homogeneity between treatment and control groups regarding pre-existing skills. 
\begin{table}[t]
\centering
\caption{PVT score differences between the values at rest and those obtained on the day of the study for the sleep-deprived group. Participants in gray background are not included in the cleaned dataset ($SD_{Cleaned}$).}
\resizebox{0.48\textwidth}{!}{
\begin{tabular}{lccccc}
\toprule
\multicolumn{1}{c}{ID} & \multicolumn{1}{c}{\begin{tabular}[c]{@{}c@{}}Performance\\Score\end{tabular}} & \multicolumn{1}{c}{\begin{tabular}[c]{@{}c@{}}Mean \\ 1/RT\end{tabular}} & \multicolumn{1}{c}{\begin{tabular}[c]{@{}c@{}}10\% Slowest\\ 1/RT\end{tabular}} & \multicolumn{1}{c}{\begin{tabular}[c]{@{}c@{}}Minor\\ Lapses\end{tabular}} & \multicolumn{1}{c}{\begin{tabular}[c]{@{}c@{}}10\%Fastest \\ RT\end{tabular}} \\ \midrule
\rowcolor[HTML]{B8B8B8} 
P1 & \textbf{-2} & 0.01 & \textbf{-0.14} & \textbf{1} & 0 \\
P2 & 5 & 0.94 & 1.42 & -6 & -21 \\
\rowcolor[HTML]{B8B8B8} 
P3 & \textbf{-4} & 0.31 & 0.35 & \textbf{1} & -30 \\
\rowcolor[HTML]{B8B8B8} 
P4 & 0 & \textbf{-0.09} & \textbf{-0.47} & 0 & \textbf{1} \\
P5 & 0 & 0.58 & 0.12 & 0 & -10 \\
P6 & 25 & 1.62 & 0.75 & -26 & -67 \\
\rowcolor[HTML]{B8B8B8} 
P7 & \textbf{-1} & \textbf{-0.16} & \textbf{-0.72} & \textbf{1} & \textbf{14} \\
\rowcolor[HTML]{B8B8B8} 
P8 & 5 & 0.34 & 1.22 & -5 & \textbf{39} \\
P9 & 0 & 0.77 & 0.05 & 0 & -74 \\
P10 & 1 & 0.26 & 0.63 & 0 & -12 \\
P11 & 3 & 1.03 & 0.88 & -3 & -66 \\
\rowcolor[HTML]{B8B8B8} 
P12 & \textbf{-2} & \textbf{-0.17} & 0.12 & \textbf{2} & \textbf{23} \\
P13 & 5 & 0.31 & 0.77 & -4 & -8 \\
P14 & 0 & 0.29 & 0.18 & 0 & -15 \\
P15 & 14 & 1.27 & 1.12 & -16 & -72 \\
P16 & 1 & 0.75 & 1.06 & -1 & -38 \\
P17 & 2 & 0.84 & 1.72 & -2 & -12 \\
P18 & 1 & 0.52 & 0.44 & -3 & -23 \\
P19 & 5 & 0.4 & 0.42 & -5 & -26 \\
\rowcolor[HTML]{B8B8B8} 
P20 & \textbf{-4} & \textbf{-0.39} & 0.41 & \textbf{3} & 0 \\
\rowcolor[HTML]{B8B8B8} 
P21 & \textbf{-2} & 0.32 & \textbf{-0.47} & \textbf{1} & -27 \\
P22 & 10 & 0.6 & 0.46 & -13 & -34 \\
P23 & 3 & 0.35 & 0.04 & -5 & -50 \\ \bottomrule
\end{tabular}
}
\label{tab:differencesPVT}
\end{table}

We performed data analysis\footnote{The analysis was carried out using R version 3.3.1.} considering both \textit{$SD_{Uncleaned}$} (from here onward also referred to as uncleaned dataset) and \textit{$SD_{Cleaned}$} (from here onward also referred to as cleaned dataset). 
In particular, we carried out the following steps:
\begin{enumerate}
\item Compute the descriptive statistics for all  the  dependent variables;
\item Use violin plots to summarize and visualize the gathered data;
\item Apply Bonferroni correction~\cite{Dunn1961} (when needed) to mitigate the family-wise error rate;
\item Test our null hypotheses using the Mann-Whitney U test~\cite{Conover:1998} due to the non-normality of the data; 
\item Study the magnitude of the differences between two groups using Cliff's $d$~\cite{KampenesDHS07} as a measure of the effect size. The confidence intervals for the effect size were also calculated to interpret its precision;
\item Provide the means percentage reduction as a less robust---though more intuitive---effect size indicator.\footnote{In particular, given $\mu_{RS}$ and $\mu_{SD}$, corresponding respectively to the mean values of the control and treatment groups for a given dependent variable, the means' percentage reduction is computed as $\frac{(\mu_{RS}-\mu_{SD})}{\mu_{RS}}\%$.} 
\end{enumerate}

\subsection{Operation}\label{sec:operation}
\label{sec:preOp}
In the following subsections, we describe the steps taken during the training and experimental operation. 

\subsubsection{Training}
The IS course was accompanied by four hands-on training sessions of three hours each (see Figure~\ref{fig:overview}). 
The sessions took place in a didactic laboratory at the University of Basilicata over a period of one month, and all the participants in our experiment attended the sessions.
During the sessions, the students improved their knowledge regarding the development of unit tests in Java using the JUnit framework, refactoring, and TFD\@.
They also familiarized themselves with the Eclipse IDE and the Besouro plug-in, later used in the experiment.
Throughout the training sessions, the participants were asked to use TFD (although it was not mandatory) to solve several code katas of increasing difficulty, and worked on home assignments to further practice the contents of the course.
All participants received the same training. 

The material used in the training session (e.g., slides, practice tasks) was the same used in~\cite{Fucci:2013eo}.
This material was translated from the English language to the Italian language to avoid that different familiarity levels of the participants with the English  would affect experimental results.  

The participants followed a five-steps procedure during each training session:
\begin{enumerate}
	\item import a starting Eclipse project containing a stub of the expected API signature for the assigned programming  task;\footnote{This was done to avoid mundane tasks, but also to have a consistent naming convention.}
	\item start the Besouro Eclipse plugin;
	\item implement a user-story card\footnote{A card is usually a very high-level definition of a requirement; it contains just enough information for the developers to reasonably estimate the effort required to implement it~\cite{williams2003xp}.} (or simply card, from here on) using TFD\label{item:4};
	\item stop the Besouro Eclipse plugin.
\end{enumerate}

Regarding \ref{item:4}), we asked the participants to implement \textit{confirmations} (i.e., conditions that need to be satisfied) for the card. 
The confirmations were of incremental difficulty, each building on the results of the previous one. 
However, we did not impose any order the participants had to follow to implement the confirmations, but we suggested them to follow the order in which they appeared.
We also did not suggest any approach to browse, run, execute, and debug source code.
That is, participants could freely use all the features present in the Eclipse IDE\@.

We had information on the composition of the experimental groups after the last training session, before the experiment, was over (left side of Figure~\ref{fig:overview}).
At the end of this session (November 30th, 2015), we administered the questionnaire to measure participants' self-perceived experience level (see Table~\ref{tbl:experience}).

\subsubsection{Experiment operation} \label{sec:expOper}
The experimental task was tackled at 9.00 a.m.\ on December 12, 2015, under controlled conditions in a laboratory at the University of Basilicata.
The experimental operation followed the same five-steps procedure of the training session.
We provided the participants with a hard copy of the task card shown in Figure~\ref{fig:card}. This card was never shown to the students during the training sessions. 

The experimental session lasted for 90 minutes, after which the participants returned source code of the solution they implemented. 
The projects were later used to extract the metrics necessary to assess the constructs.
We impose a time limit to perform the task in the experimental session because this allows a better evaluation of developers' performance~\cite{Simula.simula.2934}. 

We allowed all the participants to use the Internet to accomplish the task, but we forbid them to use the Internet to communicate with one another. Two supervisors made sure that no interaction among participants took place.

\begin{figure}[!ht]

\begin{framed}
{\scriptsize Pig Latin is a language game in which words in English are altered. The objective is to conceal the words from others not familiar with the rules. The reference to Latin is a deliberate misnomer, as it is simply a form of jargon, used only for its English connotations as a strange and foreign-sounding language. We ask the participants to implement the following confirmations:

\begin{enumerate}
\item Create a PigLatin class that is initialized with a string
\begin{itemize}
\item[-] The string is a list of words separated by spaces: ``hello world''
\item[-] The string is accessed by a method named phrase
\item[-] The string can be reset at any time without re-initializing
\item[-] PigLatin.new('hello world')
\end{itemize}

\item Create a TranslateMethod, namely a method that translates the phrase from English to pig-latin.
\begin{itemize}
\item[-]   The method will return a string.
\item[-] The empty string will return nil.
\item[-]  `` '' translates to nil
\end{itemize}

\item Translate words that start with vowels.
\begin{itemize}
\item[-] Append ``ay'' to the word if it ends in a consonant.

         example: ``ask'' translates to ``askay''
\item[-] Append ``yay'' to the word if it ends with a vowel.

          example: ``apple'' translates to ``appleyay''
\item[-] Append ``nay'' to the word if it ends with ``y''.

         example: ``any'' translates to ``anynay''
\end{itemize}

\item Translate a word that starts with a single consonant.
\begin{itemize}
\item[-] Removing the consonant from the front of the word.
\item[-] Add the consonant to the end of the word.
\item[-] Append ``ay'' to the resulting word.
         
         example: ``hello'' translates to ``ellohay''
         
         example: ``world'' translates to ``orldway''
\end{itemize}
\item Translate words that start with multiple consonants.
\begin{itemize}
\item[-] Remove all leading consonants from the front of the word.
\item[-]  Add the consonants to the end of the word.
\item[-] Append `ay' to the resulting word.

          example: ``known'' translates to ``ownknay''
         
          example: ``special'' translates to ``ecialspay''
\end{itemize}

\item Support any number of words in the phrase.

         example: ``hello world'' translates to ``ellohay orldway''
\begin{itemize}
\item[-] Each component of a hyphenated word is translated separately.
        
        example: ``well-being'' translates to ``ellway-eingbay''
\end{itemize}

\item Support capital letters.
\begin{itemize}
\item[-] If a word is capitalized, the translated word should be capitalized.

         example: ``Bill'' translates to ``Illbay''
         
         example: ``Andrew'' translates to ``Andreway''
\end{itemize}
\item Retain punctuation.
\begin{itemize}
\item[-] Punctuation marks should be retained from the source to the translated string
        
        example: ``fantastic!'' translates to ``anfasticfay!''
        
        example: ``Three things: one, two, three.'' translates to ``Eethray ingsthay: oneyay, otway, eethray.''
\end{itemize}
\end{enumerate}
}
\end{framed}
  \caption{Card administered to the participants }\label{fig:card}
\end{figure}

\subsection{Experimental object}\label{sec:tasks}
The experimental object is a programming exercise (i.e., a code kata), that consisted in implementing the PigLatin program commonly used to demonstrate TFD principles~\cite{Elkner} .
This card (Figure~\ref{fig:card}) contains eight confirmations. Each confirmation has a short description and at least one example of input and expected output.
In this sense, confirmations are a sort of acceptance tests for the card. The acceptance tests used to calculate $PAAP$ were different from the ones shown in Figure~\ref{fig:card}.
To implement the card, the participants used the Java programming language (version 6) and JUnit (version 4). 
We provided the participants with a template Eclipse project containing a stub of the expected API signature. Both groups tackled the same experimental task.

\begin{table*}[t]
\centering
\normalsize
\caption{Descriptive statistics for the considered metrics. 
Values in parentheses refer to the original dataset ($SD_{Uncleaned}$).
}
\resizebox{1\linewidth}{!}{
\begin{tabular}{lccccccccccccccc} \toprule
\multirow{2}{*}{Metric} &  \multicolumn{7}{c}{RS} & & \multicolumn{7}{c}{SD} \\
\cline{2-8} \cline{10-16} 
&  Min & Max & Mean & Q1 & Median &Q3& St.\ Dev. & & Min & Max & Mean & Q1 & Median &Q3& St.\ Dev.  \\ \midrule
 \multirow{2}{*}{PAAP} & \multirow{2}{*}{0} & \multirow{2}{*}{53.85} & \multirow{2}{*}{28.57} & \multirow{2}{*}{15.38} & \multirow{2}{*}{38.46} & \multirow{2}{*}{38.46} & \multirow{2}{*}{16.344} & & 0 & 53.85 & 14.36 & 7.692 & 15.38 & 15.38 & 14.498 \\ 
&  &  &  &  &  &  &  & & (0) & (53.85) & (15.05) & (7.692) & (15.38) & (15.38) & (13.013) \\
\midrule
 \multirow{2}{*}{\#episodes} & \multirow{2}{*}{0} & \multirow{2}{*}{16} & \multirow{2}{*}{7.571} & \multirow{2}{*}{2 }& \multirow{2}{*}{9} & \multirow{2}{*}{12} & \multirow{2}{*}{5.39} & & 0 & 13 & 4.267 & 0 & 4 & 6 & 4.59\\ 
&  &  &  &  &  &  &  & &  (0) & (13) & (4.696) & (1) & (4) & (6.5) & (4.3)  \\\midrule
\multirow{2}{*}{\%conformance} & \multirow{2}{*}{0 }& \multirow{2}{*}{100} & \multirow{2}{*}{45.43} & \multirow{2}{*}{0} & \multirow{2}{*}{50} & \multirow{2}{*}{73} & \multirow{2}{*}{37.682} & & 0 & 100 & 25.27 & 0 & 0 & 55 & 38.653 \\ 
&  &  &  &  &  &  &  & & (0) & (100) & (24.96) & (0) & (0) & (50) & (36.738)  \\
\bottomrule


\end{tabular}
}
\label{tab:descriptiveStat}
\end{table*}

\section{Results}
\label{sec:Findings}
In this section, we first present the descriptive statistics for the dependent variables (Section~\ref{sec:descriptive}) and then the results of the statistical hypothesis testing (Section~\ref{sec:statInference}). We conclude by reporting  results from  additional analyses (Section~\ref{sec:additional}). 

\begin{figure*}[ht]
{
 \subfigure[]{\label{fig:PAAP}\includegraphics[width=0.33\linewidth]{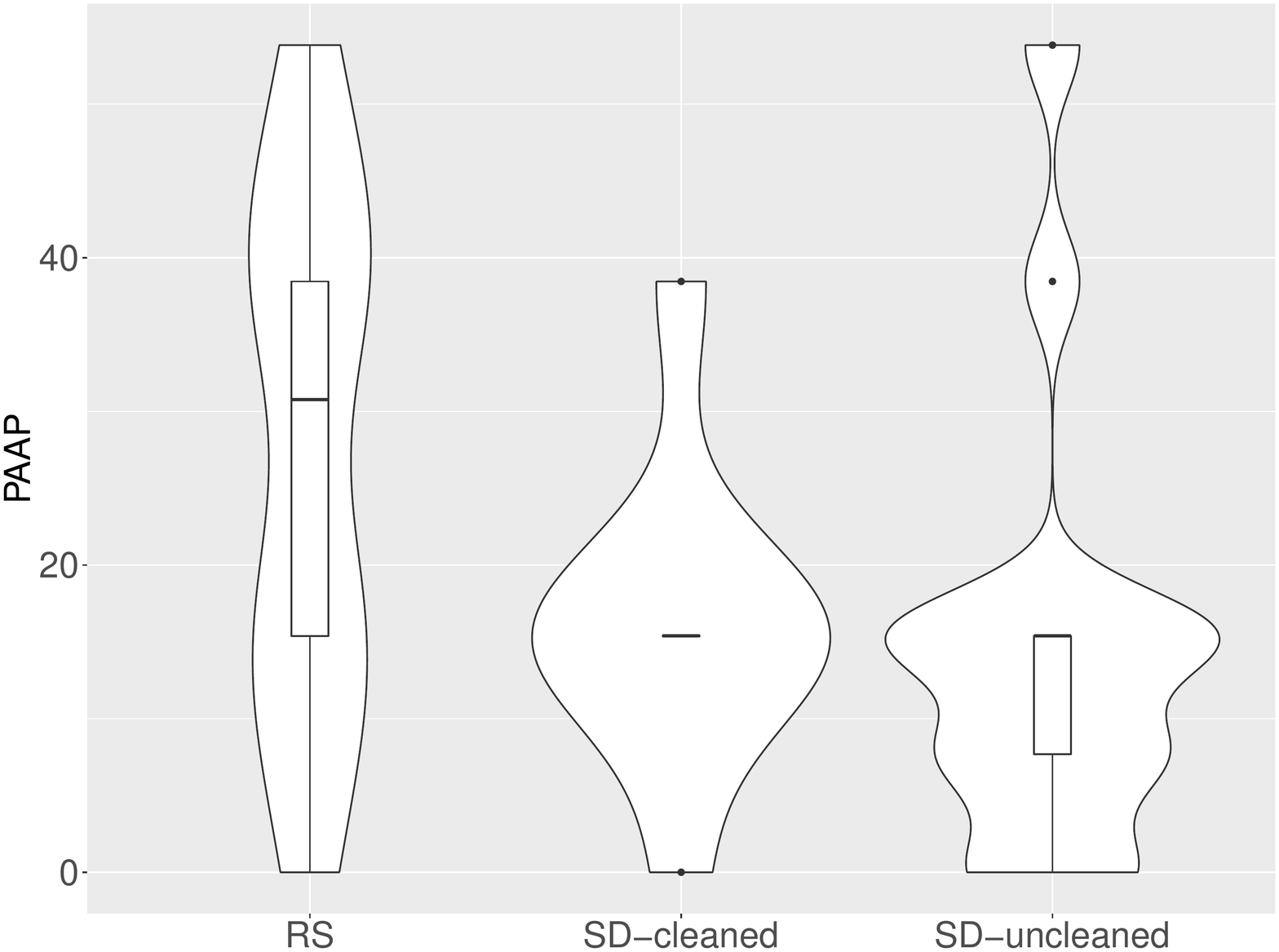}}
 \subfigure[]{\label{fig:episodes}\includegraphics[width=0.33\linewidth]{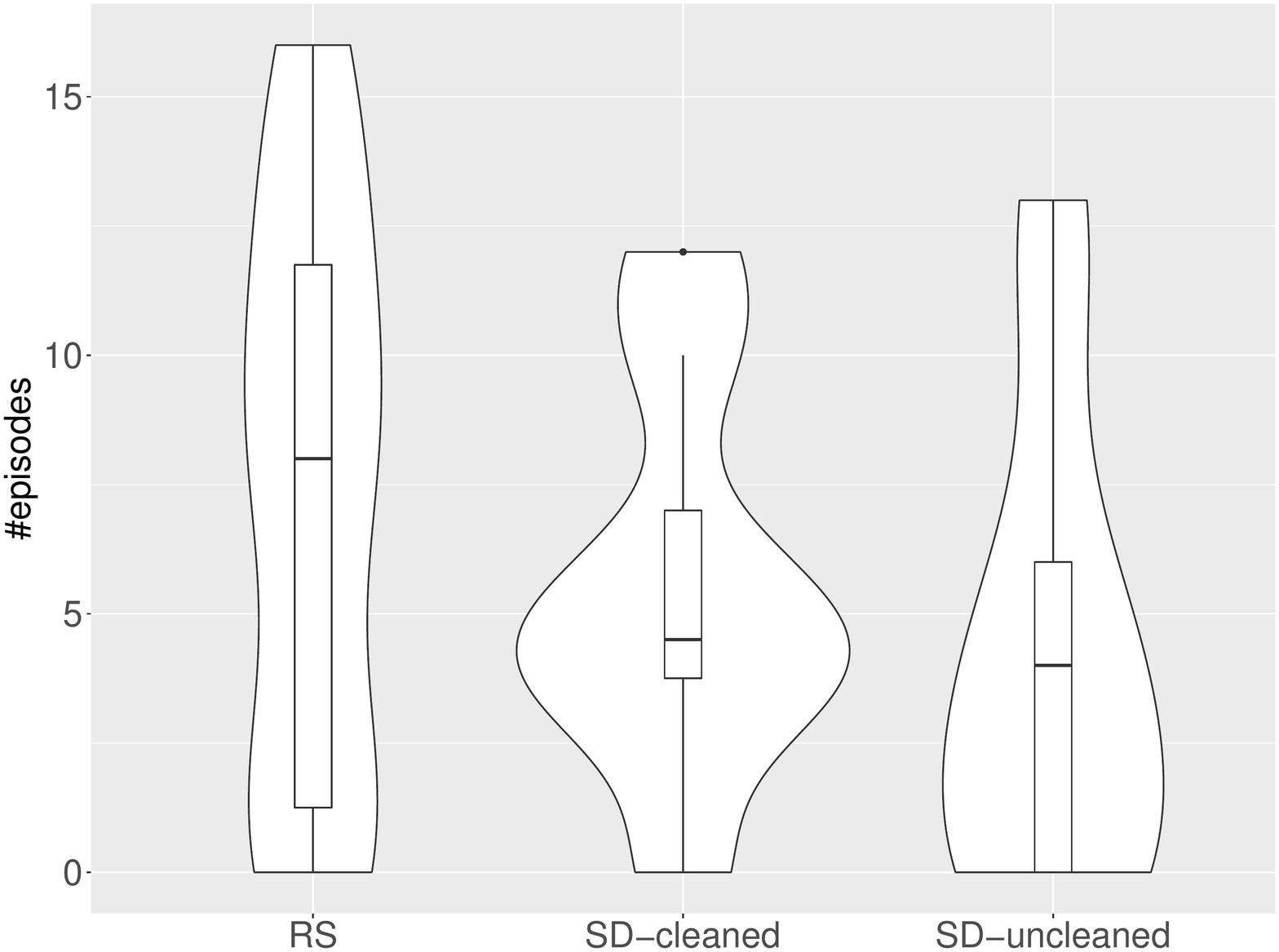}}
 \subfigure[]{\label{fig:conf}\includegraphics[width=0.33\linewidth]{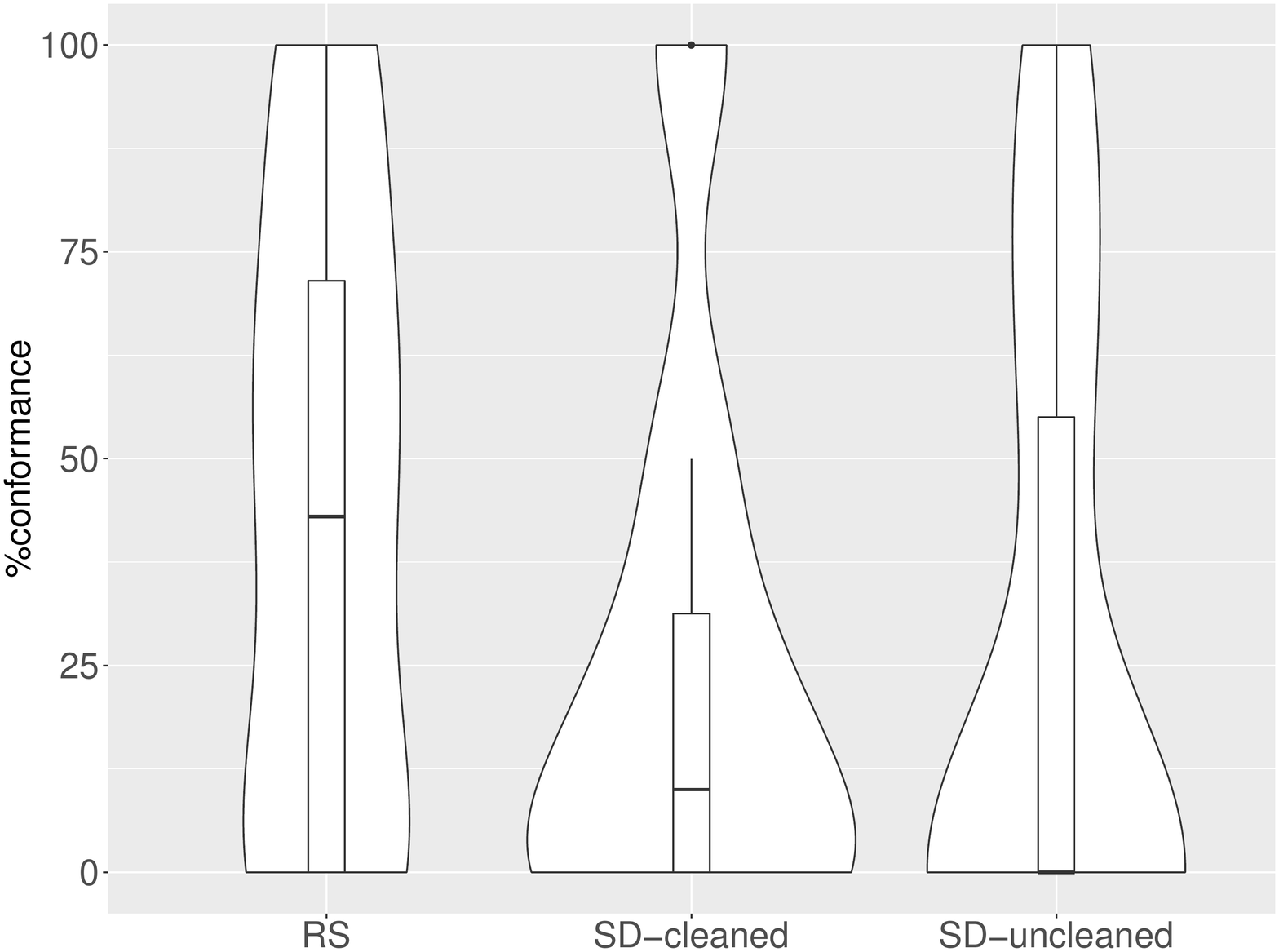}}
}
\caption{Violin-plots representing the dependent variables distribution for \subref{fig:PAAP} quality ($PAAP$),  \subref{fig:episodes} engagement ($\#episodes$),  \subref{fig:conf} test-first development conformance ($\%conformance$) for the regular ($RS$) and sleep-deprived ($SD$) groups (cleaned and uncleaned datasets).}
\label{fig:boxes}
\end{figure*}


\begin{table}[!t]
\centering
\caption{Shapiro-Wilk test results to assess the normality of dependent variable distributions ($\alpha$ = .05)}
\resizebox{1\linewidth}{!}{
\begin{tabular}{lccc}
\toprule
\multicolumn{1}{c}{DV} & \multicolumn{1}{c}{Shapiro-Wilk} & \multicolumn{1}{c}{p-value} & \multicolumn{1}{l}{Normally distributed} \\ \midrule
PAAP                                   & .48                                & 2.02e-11                    & No                                        \\
\#episodes                             & .91                                & .02                         & No                                        \\
\%conformance                          & .79                                & 2.35e-06                    & No                                        \\ \bottomrule
\end{tabular}
}
\label{tbl:shapiro}
\end{table}
\begin{table*}[!t]
\centering
\normalsize
\caption{Null-hypothesis testing, and effect size results. The $*$ indicates values significant after applying the Bonferroni correction ($\alpha=0.016$). Values in parentheses refer to the original dataset ($SD_{Uncleaned}$).}
\resizebox{0.8\linewidth}{!}{
\begin{tabular}{llccccc}
\toprule
\multicolumn{1}{c}{Research question}&\multicolumn{1}{c}{Hypothesis} & \multicolumn{1}{c}{Mann-Whitney U} & \multicolumn{1}{c}{p-value} & \multicolumn{1}{c}{effect size} & \multicolumn{1}{c}{\% reduction} & \multicolumn{1}{c}{95\% effect size CI}  \\ \midrule
RQ1&$H0_{QLTY}$             & 208 (229)           & .005* (.006*)                 & -0.498 (-0.466)                 & -49.73 (-47.32)                  & {[}-1.58, -0.20{]} ({[}-1.41, -0.20{]})                                \\
RQ2&$H0_{ACTV}$          & 167 (187)                & .035 (.040)                 & -0.359 (-0.308)                 & -43.64 (-37.97)                  & {[}-1.31, 0.04{]} ({[}-1.17, 0.01{]})                                  \\
RQ3&$H0_{PROC}$           &163 (179)            & .046 (.027)                 & -0.321 (-0.325)                 & -44.37 (-45.08)                  & {[}-1.18, 0.15{]} ({[}-1.13, 0.05{]})                                 \\ \bottomrule
\end{tabular}
}
\label{tab:dataAnalysis}
\end{table*}
%
%
%

\subsection{Descriptive statistics} \label{sec:descriptive}
Table~\ref{tab:descriptiveStat} reports the descriptive statistics for the metrics used to measure the dependent variables; the violin-plots (Figure~\ref{fig:boxes}) graphically summarize their distributions and the superimposed boxplots report the summary statistics.
On average, the $RS$ group performed twice as well as the $SD$ group regarding $PAAP$, although their variation (i.e., standard deviation) is similar.
Figure~\ref{fig:PAAP} visually shows that the  quality of the software produced by sleep-deprived developers is below that of the developers working under normal-sleep condition.
The boxplots representing the two groups do not overlap; therefore, we expect a clear difference.
The difference is less clear for $\#episodes$. Figure~\ref{fig:episodes} shows that sleep-deprived participants are less active. 
However, the observations at the bottom of the boxplots are similar.
Six participants in the $SD$ group (one in the $RS$ group) were not active---i.e., they did not accomplish any of the  development activities, yielding a $\#episode$ value of zero.
Figure~\ref{fig:conf} shows that sleep-deprived developers were less capable of following TFD, although such difference is not as remarked as for the other two dependent variables.
The participants in the $SD$ group follow the TFD process 20\% of the time less compared to those in the $RS$ group (i.e., 25\% vs. 45\% in $\%conformance$ score); however, there is a considerable variation in both groups.
In the $SD$ group, 12 participants did not follow TFD (seven in the $RS$ group)---i.e., these participants had a $\%conformance$ score equals to zero (see Appendix~\ref{app:2} for a post-mortem analysis of this result).
The results are similar when all the original participants are considered in the $SD$ group---i.e., no data cleaning based on PVT scores was applied. 
In general, Figure~\ref{fig:boxes} shows that the distributions of the dependent variables are more uniform for $RS$, whereas they tend to be skewed towards the bottom (i.e., lower values) for $SD$. 

\subsection{Statistical inference} \label{sec:statInference}
We applied Bonferroni correction when testing the hypotheses. Therefore, the $\alpha$ considered as a threshold to reject the null hypotheses was .016 (i.e., the standard $\alpha$ .05 divided by the three hypotheses being tested).
We used the non parametric Mann-Whitney U test because the dependent variables are not normally distributed according to the result of Shapiro-Wilk test~\cite{shapiro65}, reported in Table~\ref{tbl:shapiro}.

In this section, we discuss the results based on Table~\ref{tab:dataAnalysis} which reports the test statistics, p-values, and effect sizes. 

\noindent\textbf{RQ1:  PAAP}.
The test result allowed us to reject the null hypothesis $H0_{QLTY}$, the sleep-deprived group ($SD$) performs worse than regular sleep group ($RS$) regarding the percentage of assert passed ($PAAP$).
The estimated effect size is to be considered \textit{medium}, and its negative sign is consistent with the direction of the alternative hypothesis. 
As the null hypothesis is rejected, the confidence interval of the effect size does not include zero, but its range is quite wide. 
In other words, an effect of sleep deprivation on external software quality exists and it is showed to be medium in our quasi-experiment. However, a more precise estimation of its real size, which could be as large as 1.58, requires further replications.
From a practical perspective, sleep deprivation can deteriorate software quality, measured through $PAAP$, of about 50\% as indicated by the percentage reduction estimator.

\noindent\textbf{RQ2: ACTV}.
The null hypothesis about developers' engagement ($H0_{ACTV}$) could not be rejected at the considered p-value threshold.
Although the medium effect size is consistent with the direction postulated in the alternative hypothesis, its wide confidence interval includes zero.
Therefore, we cannot be sure that an effect of sleep on developers' engagement exists in reality, but if it does, it would likely be detrimental (as large as 1.31).
Considering that, sleep-deprived developers were less engaged with the task---i.e., perform 43\% fewer development activities with respect to developers working under normal-sleep condition.

\noindent\textbf{RQ3: PROC}.
The null hypothesis regarding developers' adherence to the TFD process ($H0_{PROC}$) could not be rejected at the considered p-value threshold.
The estimated medium effect size is negative as postulated by the alternative hypothesis. 
As for the case of $\#episodes$, given the evidence collected in this experiment, we cannot be sure whether an effect of sleep deprivation exists in reality for $\%conformance$. 
The result suggests that the effect is likely to be negative and as large as 1.18.
From the effect-size estimate in this study, sleep-deprived software developers perform TFD 44\% less compared to software developers under normal-sleep condition.

\subsection{Additional results}
\label{sec:additional}
In this section, we report the results of additional analyses which were not originally planned when the experiment was designed, but emerged only after we looked at the data.
We report the additional hypotheses and results in a separate section to mitigate researcher bias~\cite{john2012measuring}.

\noindent \textbf{Differences in source code edits}. 
The data collected from the participants' IDE allowed us to gather further insights about their behavior. 
We conjecture that the participants in the $SD$ group are likely to commit syntactical mistakes (e.g., illegal sequence of tokens) when writing code due to their susceptibility to distractions~\cite{Mark:2016}.
Such mistakes need to be fixed by editing the source code (e.g., by renaming or removing tokens) so that the project can be compiled and the unit tests can be correctly executed.
Therefore, from the log files registered by the Besouro Eclipse plugin, we gathered the editing actions\footnote{Actions are the basic interactions within the IDE which together form the development activities presented in Table~\ref{tbl:types}} that involved deleting or renaming identifiers (e.g., variable and method names) which fix syntactical mistakes.
We define $FIX$ as the ratio of such actions over the total amount of actions registered.

We formulated the following one-tailed null hypothesis $H0_{FIX}$ to check the effect of sleep deprivation on the amount of fixing actions.   
\begin{itemize}
    \item \textit{$H0_{FIX}$}: The amount of fixing actions performed by developers who stay awake the night before an implementation task is not larger than the amount of fixing actions performed by developers who slept normally. 
\end{itemize}
The alternative hypothesis follows:
\begin{itemize}
    \item \textit{$H1_{FIX}$}: The amount of fixing actions performed by developers who stay awake the night before an implementation task is larger that the amount of fixing actions performed by developers who slept normally. 
\end{itemize}

For the analysis, we used the same steps presented in Section~\ref{sec:AnalysisProc}.
Using the Shapiro-Wilk test (W = .941, p-value = .023), we could not show that the $FIX$ is normally distributed; therefore, we compared the two groups in terms of $FIX$ using the same non-parametric test used for the previous hypotheses, i.e., the Mann-Whitney U test. 


It appears that sleep-deprived participants perform more fix actions than the ones at rest ($W$ = 82.5,  $p-value$ = .008). 
The effect size is .41 , CI = [.2, .46], and mean percentage reduction is 54\%.
The \textit{medium} effect size and its confidence interval allow us to conclude that sleep-deprived developers tend to do more fixing actions (approximately 54\%)  than developers who slept normally. 
Our explanation is that sleep deprivation has an effect on driving changes to the source code due to inattention.

\noindent \textbf{Correlations with waking hours.}
In the pre-questionnaire, we asked the participants to report the number of waking hours before performing the experimental task.   
Considering the $SD$ group, the average number of awake hours ($AWAKE$) was 20.73 (median = 18, sd = 4.31, range = [16, 28]).
The result of a Shapiro-Wilk test shows that the variable is not normally distributed ($W$ = 0.803, $p-value$ = .004). 
Subsequently, we assess the correlations between $AWAKE$ and the three dependent variables of our study using Spearman correlation coefficient, deemed appropriate for non-normally distributed variables.
In Table~\ref{tbl:corrs}, we report the correlation analysis results.  
\begin{table}[!t]
    \centering
    \caption{Sprearman correlations between the number of waking hours ($AWAKE$) and the three dependent variables.}
\label{tbl:corrs}
    \begin{tabular}{lcc}
        \toprule
                    & Spearman $\rho$ & p-value \\ \midrule
    $PAAP$          & -.382           & .151    \\
    $\#episodes$    & -.345           & .206    \\
    $\%conformance$ & -.108           & .700    \\ \bottomrule
    \end{tabular}
    \end{table}

Although the Spearman coefficients show a negative correlation---i.e., the longer a participant is awake, the less she performs---none was statistically significant. 
Nevertheless, this preliminary outcome can be used to create a baseline for future experiments---e.g., when deciding the different levels of sleep deprivation in further studies.

\noindent \textbf{Perceptions of sleep-deprived participants.}
In a post-questionnaire, we gathered open-form feedback from the participants after the completion of the experimental task.
In particular, three participants in the $SD$ elaborated on the experience of being sleep deprived and the perceived effect of sleep deprivation on their performance during the task\footnote{The quotations were originally recorded in Italian and translated into English.}.

\textit{P2} - \say{It is difficult to undertake such task without any sleep. I was lacking focus, attention, and the ability of reading and understanding properly.}

\textit{P6} - \say{In my opinion, under a sleep deprivation condition you quickly lose focus. Like in my case, I easily had doubts about pieces of code which I would otherwise grasp.}

\textit{P17} - \say{Sleepiness remarkably slowed me down in each aspect, especially when it came to logical thinking. The task problem is surely simple but it was so difficult for me to think under a sleep deprivation condition.}

It appears that sleep-deprived participants experience loss of focus and attention.
From P6, it appears that sleep deprivation provokes a loss in self-confidence and causes uncertainties in the participant's.
Further studies can survey this effect and try to quantify to which extents it applies to software development.

\section{Discussion}\label{sec:Discussion}
In the following subsections, we answer the research questions  (Section~\ref{sec:RQs}) and discuss limitations to be considered when interpreting our results (Section~\ref{sec:threats}). We also delineate practical implications and future directions for research (Section~\ref{sec:implications}). We conclude showing some lessons we learned from the execution of our experiment (Section~\ref{sec:lessonLearned}). 

\subsection{Answers to the research questions}
\label{sec:RQs}
\begin{itemize}
\item[-] \textit{\textbf{RQ1}. Does sleep deprivation decrease the quality of the solution to a programming task?}

Since we were able to reject  $H0_{QLTY}$ we answer RQ1 as follows: \textbf{developers who forewent one night of sleep write code which is approximately 50\% more likely to not fulfill the specification with respect to the code written by developers under normal sleep condition.}
This outcome is the most important result of our study, and it is in line with similar studies, in other disciplines, dealing with measuring the impact of sleep on performance with other types of cognitive tasks~\cite{Walker:2009iy,3956626}. 
The answer to RQ1 supports the theory that sleep deprivation has detrimental effects on the quality (i.e., functional correctness) of software developed by novices.
In our additional analysis, we showed that sleep-deprived developers are more prone to perform editing actions to address syntax issues and, from preliminary qualitative evidence, that even simple operations can become difficult under such condition. 
It is perhaps not overly surprising, but evidence needs to be obtained through empirical studies to move from opinions and common sense to facts (e.g.,~\cite{kitchenham02preliminary,basili99,ShullCVJ08}), as well as to have a first understanding of the size of the impact~\cite{sullivan2012using} that sleep deprivation can have on software development activities.

\item[-] \textit{\textbf{RQ2}. Does sleep deprivation decrease the developers' activeness in writing source code?}
Although the results are not conclusive, we have evidence that one night of sleep deprivation can be harmful to the developers' activeness, with a loss of about 43\% reported in this study. 

\item[-] \textit{\textbf{RQ3}. Does sleep deprivation decrease the ability of developers to follow the TFD process?}
In this study, we could not conclude that the ability to apply TFD is impacted by sleep.
However, the evidence shows that sleep-deprived developers can encounter some difficulties applying TFD\@.
\end{itemize}

The main take-away from our quasi-experiment  can be summarized as follows:
\begin{framed}
\noindent
One night of sleep deprivation is detrimental for software developers.
In particular, sleep-deprived developers produce software of lower quality (i.e., functional correctness).
\end{framed}

To increase our confidence in the results, it would be advisable to replicate this study with a design that takes into consideration individual characteristics of the participants, such as gender and age. 
It would be also important to conduct replications with developers with different levels of programming experience (i.e., more/less experienced) over multiple days with less/no sleep over a longer duration (e.g., one working week), or at different levels of sleep deprivation.
In such regard, our additional results showed that (relatively) short awake-time differences among participants in the SD group  (approximately four hours in the case of our experiment) may not be enough to show differences in performance. 
Based on this preliminary results, we suggest that a sensible second level of sleep deprivation condition, under which software developers can be studied, is between 25 and 28 hours of awake time~\cite{DINGES:2004tl}.

\subsection{Threats to validity}\label{sec:threats}
Sleep-deprivation studies are laborious and expensive to carry out leading to compromises in the design~\cite{alhola2007sleep}. 
We discuss the threats that can affect the results following the recommendations in Wohlin \textit{et al.}~\cite{wohlin12}. 
We ranked the validity threats from the most to the least relevant for this study. 
In particular, as we are testing a theory regarding sleep in the context of software engineering, we prioritize internal validity (i.e., warrant that a causal relationship exists) and construct validity (i.e., the metrics and instruments represent the constructs specified in the theory) rather than external validity (i.e., the generalization of the results to a context wider than that of this study).

\subsubsection{Internal validity}
The lack of randomization is the hallmark of quasi-experiment, and it is common practice in physiology studies where several treatments cannot be allowed to subjects (e.g.,~\cite{shadish2002experimental}). 
In our case, this can have an impact on the causal relationship between sleep and the dependent variables. 
For instance, the participants who are more accustomed to sleeping for a shorter amount of time may have opted to be included in the $SD$ group (i.e., compensatory rivalry).
The lack of randomization also prevented the possibility to apply blocking. For instance, the participants' age and gender can impact the causal relation between sleep deprivation and the response variables. 
However, after the participants decided to be included in one experimental group, we observed that the $SD$ group was approximately two years older than the $RS$ group.  
This difference should not affect the outcomes as the natural sleep-awake cycle is not altered for subjects in the age range of the participants~\cite{alhola2007sleep}.
According to the available medical evidence, such alteration becomes apparent with aging (i.e., in subject older of approximately 55 years)~\cite{alhola2007sleep,philip2004age}.

The \textit{selection} threat of letting volunteers take part in the study can influence the results (45 out of 95 students enrolled in the IS course) as the sample can include participants with specific characteristics, for instance, more motivated developers. 
To deal with this kind of threat, we administered a questionnaire  at the beginning of the course to assess the homogeneity of  participants regarding the relevant pre-existing skills that might impact the study results. 
We did not find any statistically significant difference between the $RS$ and $SD$ groups regarding their knowledge of Java programming, unit testing, TFD, and Eclipse IDE.  
Similarly, we did not observe differences among the two groups when analyzing the assignment performed before the experiment took place (see Appendix~\ref{app:1}).  

\subsubsection{Construct validity}
In this study, the adherence to treatment can be problematic. 
We measured the independent variable, \textit{sleep},  not only through self-assessment, but also using a standard test from the clinical literature, namely the PVT test~\cite{dinges1985microcomputer}. 

The dependent variables may suffer from the \textit{mono-method bias}---i.e., a single type of measure is used to assess each construct.
However, we exploited metrics existing in the literature (e.g.,~\cite{Hofer:2009kd,erdogmus2005,madeyski2009test}), thus strengthening their reliability.
The metrics were calculated automatically and were less prone to human measurement errors and rater subjectivity.

The participants were aware of being part of a study regarding the effects of sleep deprivation on software development practices; thus, posing the risk of \textit{hypotheses guessing}. 
However, since the participants were not aware of the specific constructs being investigated, this threat can be considered addressed.

\subsubsection{Conclusion validity}
We addressed the threat to conclusion validity by using robust statistical methods and carefully checking the assumptions of the statistical tests used in the analysis.
Moreover, we controlled for the error rate by applying Bonferroni correction when inferring the results. 

Another threat to the conclusion validity is the \textit{reliability of treatment implementation}. 
In our case, we could not observe the sleep treatment directly---i.e., we could not require the participants to be present in a controlled environment (e.g., a laboratory) where we could observe them sleeping (or not) during the night before the experimental task. 
However, rather than relying on self-assessment, we adopted PVT as it is the commonly used test in sleep research.

\subsubsection{External validity}  
We sampled the participants by convenience from a population of students at the University of Basilicata.
Therefore, generalizing the results to a different population (e.g., professional software developers) might pose a threat of \textit{interaction of selection and treatment}. 
Our experiment involved basic programming skills; therefore, using of students as participants allow us to obtain reliable results~\cite{Feitelson15}. 
Using students participants brings various advantages, such as homogeneous prior knowledge, the availability of a large sample~\cite{Verelst:2004}, and the possibility to cheaply test experimental design and initial hypotheses~\cite{Falessi:2017bg}.
Correspondingly, a threat of \textit{interaction of setting and treatment} exists due to the non-real-world experimental task used. 
This would equally affect the results of the participants in the participants in both groups. 
That is, if there is a difference on a non-real-world task, we could speculate that this difference could increase in case of more complex development tasks.  
The settings are made more realistic by including TFD, a software development process used in industrial settings.

\subsection{Implications and future extensions} 
\label{sec:implications}
In this section, we delineate how the results of our quasi-experiment fit in the emerging area of software engineering that focuses on physiological aspects.  
\begin{itemize}
\item We complement the results from the work by M{\"u}ller and Fritz~\cite{muller2016using} by showing that sleep deprivation---usually associated with reduced blood flow in several regions of the brain and changes in body temperature~\cite{thomas2000neural}---causes a severe decrease in the capacity of developers to write code that matches functional requirements.
Medical research has shown that sleep deprivation negatively affects the responsiveness to stimuli from the same kind of emotions~\cite{pilcher2015sleep}.
These outcomes can explain our results regarding sleep-deprived developers low engagement in performing even a small development task.

\item We provided some evidence supporting the idea that sleep deprivation can hinder the application of TFD\@.
In fact, to be adequately applied, TFD requires discipline and rigor~\cite{martin2007professionalism, jeffries2007guest}.
The detrimental effects of sleep deprivation on the developers' attention level~\cite{alhola2007sleep} may explain our results.
The perception of the participants in our study, reported in the post-questionnaire, are in line with the results of Sarkar and Parnin's~\cite{sarkar2017characterizing} work investigating fatigue. 
Fatigue, as a result of sleep deprivation, caused participants to lose focus and hindered them from thinking logically even for a simple problem---both effects are also reported by Sarkar and Parnin's survey.
The lack of focus was manifested in the larger number of syntax fix the sleep deprived participant needed to perform. 
Sleep deprivation can be detrimental for development practices and software engineering tasks for which developers' attention level is crucial. This point deserves further empirical  investigations and it is relevant for practitioners and researchers.

\item Sleep deprivation is a phenomenon that commonly occurs in software development, for example when deadlines are approaching~\cite{claes2017abnormal,chilton2005person,mantyla2016mining}. 
Consequently, the result of this research can be exploited to inform practitioners about the adverse effects of sleep deprivation from a technical perspective.
Although our results apply to situations of total sleep deprivation, medical evidence shows similar (if not worse) results after a few nights of partial sleep restriction---e.g., less than five hours of sleep over several nights~\cite{dinges1997cumulative}. 
Sleep-deprived software developers should be aware that they are likely to produce buggy code and that such condition is likely to affect their programming performance.
Therefore, information regarding the quality of their sleep (for example, gathered using PVT) can be utilized, in addition to the other physiological and biological metrics proposed recently (e.g.,~\cite{fritz2014using, muller2015stuck}), to better support them.
Alongside, fitness trackers could be used to measure the quality of developers' sleep supporting or replacing PVT and self-assessment.
How and whether this kind of devices could help the research delineated in this paper will be the subject of our future work. 

\item Results seem to suggest that no sleep in a night reduces the quality of work, measured as functional correctness, of novice developers (i.e., undergraduate students).
Mark \textit{et al.}~\cite{Mark:2016} conducted an \textit{in-situ} study with students (76 undergraduates: 34 males and 42 females) on  how different sleep duration can affect the use of information technology. 
The authors observed that students with less sleep: \textit{i)}~have significantly shorter focus suggesting higher multitasking, \textit{ii)}~may seek out activities requiring less attentional resources, and \textit{iii)}~tend to have a bad mood and use social media more than usual. 
On the basis of these results, we can speculate that to limit the detrimental effect of sleep deprivation, the daily work of a novice developer should be customized to adapt to her sleep-wake pattern---for example, assigning her less demanding or critical tasks.
Further studies can look at how novice developers plan their tasks---e.g., decide whether more resources are needed when reviewing code implemented by a sleep-deprived developer. 
%


\item Future work should initially replicate this study to address the limitations presented in Section~\ref{sec:threats}. 
As the medical evidence shows that different sleep deprivation time leads to various responses in the participants~\cite{dinges1997cumulative}, a first improvement over the current design is to have treatments groups at different levels of sleep deprivation.
Likewise, the cumulative effects of sleep deprivation can be assessed. 
These future directions need specific experimental infrastructures that are often not available in software engineering laboratories.
For example, the effect of partial sleep restriction can be evaluated in a laboratory setting in which experimenters control that developers sleep for the right amount of time~\cite{alhola2007sleep}. 
Further replications of our study could involve researchers from fields such as medicine---specifically, somnipathy---where sleep laboratories are commonplace.
Other directions for further work include the study of the motivations behind sleep deprivation in software development, under which circumstances such condition happens (e.g.,~\cite{claes2017abnormal}), and how it is perceived in different software engineering phases. 

Our agenda includes the evaluation of \textit{i)}~what types of mistakes are more likely to be made by sleep-deprived developers; \textit{ii)}~how long developers should sleep to avoid the adverse impact of sleep deprivation on their performances taking into account their physiological traits; \textit{iii)}~the effect of sleep deprivation on other software engineering activities and practices (e.g., software maintenance and requirements elicitation).
\end{itemize}
\subsection{Lessons learned} \label{sec:lessonLearned}
Software engineering experiments involving humans are quite common~\cite{Sjoberg:TSE:2005}, however, ethical concerns in participants selection and allocation to experimental groups do not involve health-related issues which are commonplace in medicine. 
In our case, we not only had to follow a voluntary-based selection of the experimental sample as commonly happens in software engineering~\cite{Sjoberg:TSE:2005}, particularly with student participants, but also for groups allocation. 
In medical studies, and in particular for sleep deprivation studies, this is the only available strategy due to ethical reasons~\cite{alhola2007sleep}.
From our experience, this aspect:
\begin{itemize}
    \item makes it is costly to implement a dry-run of the study to assess the appropriateness of constructs, tasks, and  experimental design. To avoid \textit{wasting} resources (e.g., participants), we recommend using constructs which have been previously validated in studies targeting the same response variable(s), experimental objects, and measurement tools. 
    
    \item makes it difficult to check for treatment conformance (i.e., assuring that participants were actually sleep-deprived) compared to traditional software engineering experiment in which the treatment is applied over a short period of time and can be automatically checked or enforced. In our case, we use PVT which does not require special equipment but does not scale well, in terms of time, once the pool of participants to be tested becomes large---for a  single experimenter, testing a sample of 20 participants will take approximately four hours. Traditionally, in medical sleep deprivation studies, participants are observed in special chambers, implying that the experimenter(s) should invest the same time necessary to implement the treatment just to check that it is correctly followed. The cost of checking the conformance to treatment could be reduced using actimetry devices which are nowadays embedded in smart watches and health trackers.
    
    \item imposes a delicate tradeoff in the selection of an experimental design. We decided to employ a between-subjects design by verifying that participants in the experiment would be as homogeneous as possible in terms of existing skills relevant for the implementation of the experimental task. We excluded a within-subject design, although this seems to be the ordinary choice for medical sleep deprivation experiments. In contrast to software engineering, medical experiments focus on factors such as chemical reactions associated with sleep conditions (e.g., brain receptors \cite{goel2009neurocognitive}), self-reported measure about psychiatric disorders (e.g., anxiety~\cite{rosekind2010cost}), and cognitive tasks (e.g., attention span and learning rate~\cite{linde1992effect}). 
    Such constructs are not affected (or are to a less extent) by carryover and learning effects, typical of software engineering tasks, as they do not require specific skills (e.g., Java programming). Although repeated measure designs can be afforded in medicine, software engineering experiments emphasize tasks to observe a construct. Until a nomenclature of software engineering tasks---allowing researcher to compare them, for example, in terms of difficulty---will not available, we recommend a between-subjects design.
\end{itemize}



\section{Final Remarks}\label{sec:FinalRemarks}
In this paper, we presented the results of a first investigation about the effects of one night of sleep loss on the performance of software developers. 
We have asked the participants to implement a small Java application using the test-first development practice. One group of participants did not sleep the day before the experiment.
The results indicated that sleep deprivation has a negative effect on the capacity of software developers to produce a software solution that meets given requirements. In particular, developers  who  forewent  one  night  of sleep  write  code which  is  approximately  50\% more likely not to fulfill the functional requirements with respect to the code produced by developers under normal sleep condition. 
We observed that sleep deprivation decreases developers' engagement with the development task and hinders their ability to apply the test-first development practice. For example, we  observed a difference  of  about  44\% in the engagement with the task of  developers  who  forewent  one  night  of sleep  and developers under normal sleep condition. 
Moreover, the results showed that sleep-deprived developers performed 54\% more fixing addressing syntactic mistakes compared to developers who slept regularly. 


Our investigation i) has implications for education, research, and practice particularly when functional correctness is relevant---e.g., it might be useless to ask students, experiment participants, and developers to implement not-trivial task as their performances could be negatively affected in case of sleep deprivation. This paper adds to the new research direction in software engineering that focuses on programmer’s performance and the forces that impact it from a physiological perspective; ii) it represents a starting point for improving researchers and practitioners understanding of how software quality can benefit from monitoring developer's physiology (e.g., devising corrective actions to avoid that quality decreases due to sleep-deprived developers); and iii) provides a stepping store for follow-up studies in industry with different samples and using different kinds of empirical investigations (e.g., case studies)

Given the results of this study, we have reasons to believe that the community interested in assessing physiological factors for software developers performance should consider sleep quality in their research.
\section*{Acknowledgment}
We would like to thank the participants in our study, especially those in the sleep-deprivation group. We would also like to thank Angelo Mecca for his precious support with PVT.



%
\balance
\bibliographystyle{IEEEtran}
\bibliography{TSE}

\newpage

\appendices
\section{Analysis of participants' homogeneity}
\label{app:1}

We collected the participant's perceived skills using a questionnaire administered before the experiment session. 
As the items were rated using a Likert-scale, we used the Kolgorov-Smirnov (K-S) test to identify differences between the RS and SD groups. 
It is a test of goodness of fit for the univariate case when the scale of measurement is ordinal. 
The defined statistical hypotheses are:
\begin{itemize}
    \item$H0_{SKILL}$: There is no skill difference between sleep-deprived developers and developers who sleep regularly.
    \item$H1_{SKILL}$: There is skill difference between sleep-deprived developers and developers who sleep regularly ($SKILL_{SD} \ne SKILL_{RS}$). 
\end{itemize}
Where $SKILL$ is one of the Likert items measuring perceived experience with general programming, object-oriented programming, unit testing, test-first development, use of  Eclipse. 
The test results are reported in Table~\ref{tbl:ks}.

\begin{figure}[b]
  \centering
   \includegraphics[width=0.9\linewidth]{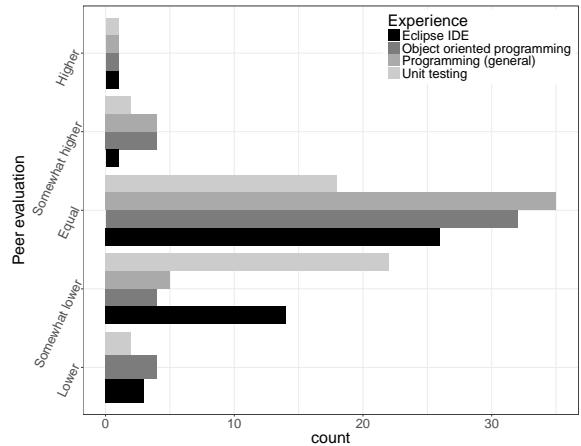}
  \caption{Participants' perception of their experience with respect to their peers.}
  \label{fig:peers}
\end{figure}

Moreover, we asked the participants to evaluate their perceived experience with respect to the rest of their classmates. 
The summary of the answers is reported in Figure~\ref{fig:peers}.

We assessed the differences between the two groups with respect to the dependent variables of this study using the score obtained from the homework assignment they carried out before the experimental task.
The task required the implementation of a formula to calculate the distance between two points on a geoid.
It was completed over a varying time span (20 hours $\pm$ 5 hours).
Table~\ref{tab:descriptiveStat2} reports the descriptive statistics for the dependent variable of this study calculated using the homework dataset. 

Considering $H0P$, the same set of null hypotheses presented in Section~\ref{sec:hypotheses} tested using the new dataset, none was rejected except for $H0P_{PROC}$ (see Table~\ref{tbl:musicphone}).

\begin{table}[b]
    \centering
    \caption{Results of the Kolgorov-Smirnoff (K-S) test comparing subjects in the sleep-deprived ($SD$) and regular sleep ($RS$) groups in terms of experience with the relevant skills. The results for the cleaned dataset ($SD_{Cleaned}$)  are reported in parentheses.}
    \begin{tabular}{@{}lcc@{}}
    \toprule
    \textbf{Experience}                  & \textbf{K-S test}    & \textbf{p-value}    \\ \midrule
    Programming (general)       & .044 (.029) & .832 (.863) \\
    Object oriented programming & .036 (.003) & .849 (.954) \\
    Unit testing                & .178 (.027) & .672 (.867) \\
    Test-first development      & .388 (.947) & .533 (.330) \\
    Eclipse IDE                 & .206 (.057) & .649 (.810) \\ \bottomrule
    \end{tabular}
    \label{tbl:ks}
    \end{table}

\begin{table}[!b]
    \centering
    \normalsize
    \caption{Null-hypothesis test results ($\alpha=0.016$).} 
    \resizebox{0.6\linewidth}{!}{
    \begin{tabular}{lcc}
    \toprule
    \multicolumn{1}{c}{Hypothesis} & \multicolumn{1}{c}{Mann-Whitney U} & \multicolumn{1}{c}{p-value}\\ \midrule
    $H0P_{QLTY}$             & 176.5           & .30   \\
    $H0P_{ACTV}$          & 167               & .036    \\
    $H0P_{PROC}$           & 129             & .013   \\ 
    \bottomrule
    \end{tabular}
    }
    \label{tbl:musicphone}
\end{table}

\begin{table*}[]
    \centering
    \normalsize
    \caption{Descriptive statistics for the dependent variable measured using the home assignment dataset.}
    \resizebox{0.7\linewidth}{!}{
    \begin{tabular}{lccccccccccccccc} \toprule
    \multirow{2}{*}{Metric} &  \multicolumn{7}{c}{$RS$} & & \multicolumn{7}{c}{$SD$} \\
    \cline{2-8} \cline{10-16} 
    &  Min & Max & Mean & Q1 & Median &Q3& St.\ Dev. & & Min & Max & Mean & Q1 & Median &Q3& St.\ Dev.  \\ \midrule
    $PAAP$ & 0 & 100 & 76.8 & 75.0 & 81.2 & 87.5 & 21.3 & & 0 & 100 & 70.4 & 61.5 & 75.0 & 87.5 & 22.6 \\ 
    \midrule
     $\#episodes$ & 0 & 17 & 6.84 & 4 & 5 & 7 & 5.12 & & 0 & 51 & 10.5 & 2.50 & 11 & 12.8 & 11.0\\ 
    \midrule
    $\%conformance$ &0 &67 & 22.9 &13 &20 & 31 &18.1 & & 0 & 100 & 10.1 & 0 & 0 & 7.5 & 23.1 \\
    \bottomrule
    \end{tabular}
    }
\label{tab:descriptiveStat2}
 \end{table*}
\section{Participants answers to post-questionnaire regarding TFD}
\label{app:2}
\begin{figure*}[ht]
{
 \subfigure[]{\label{fig:TFD}\includegraphics[width=0.45\linewidth]{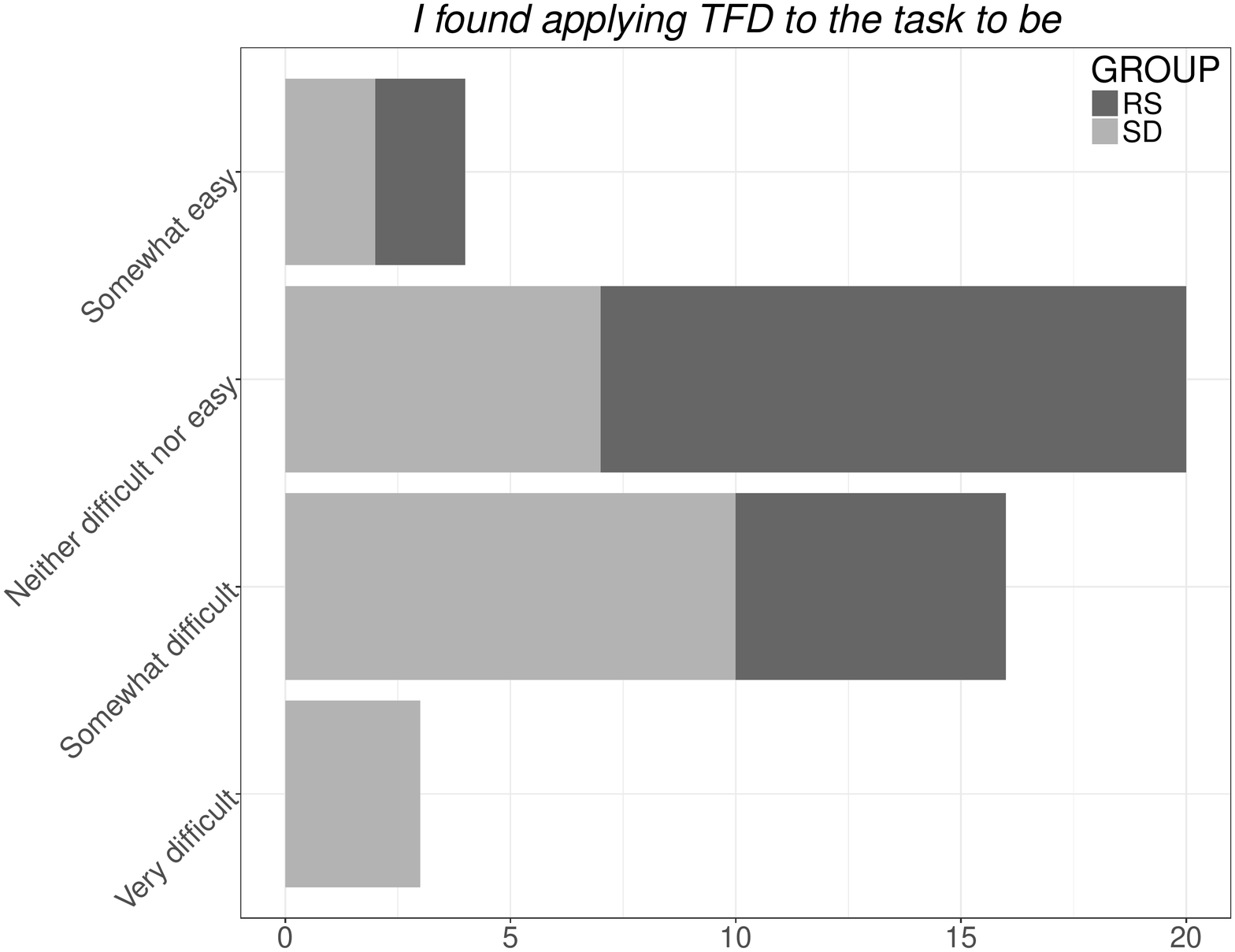}}
 \subfigure[]{\label{fig:apprach}\includegraphics[width=0.45\linewidth]{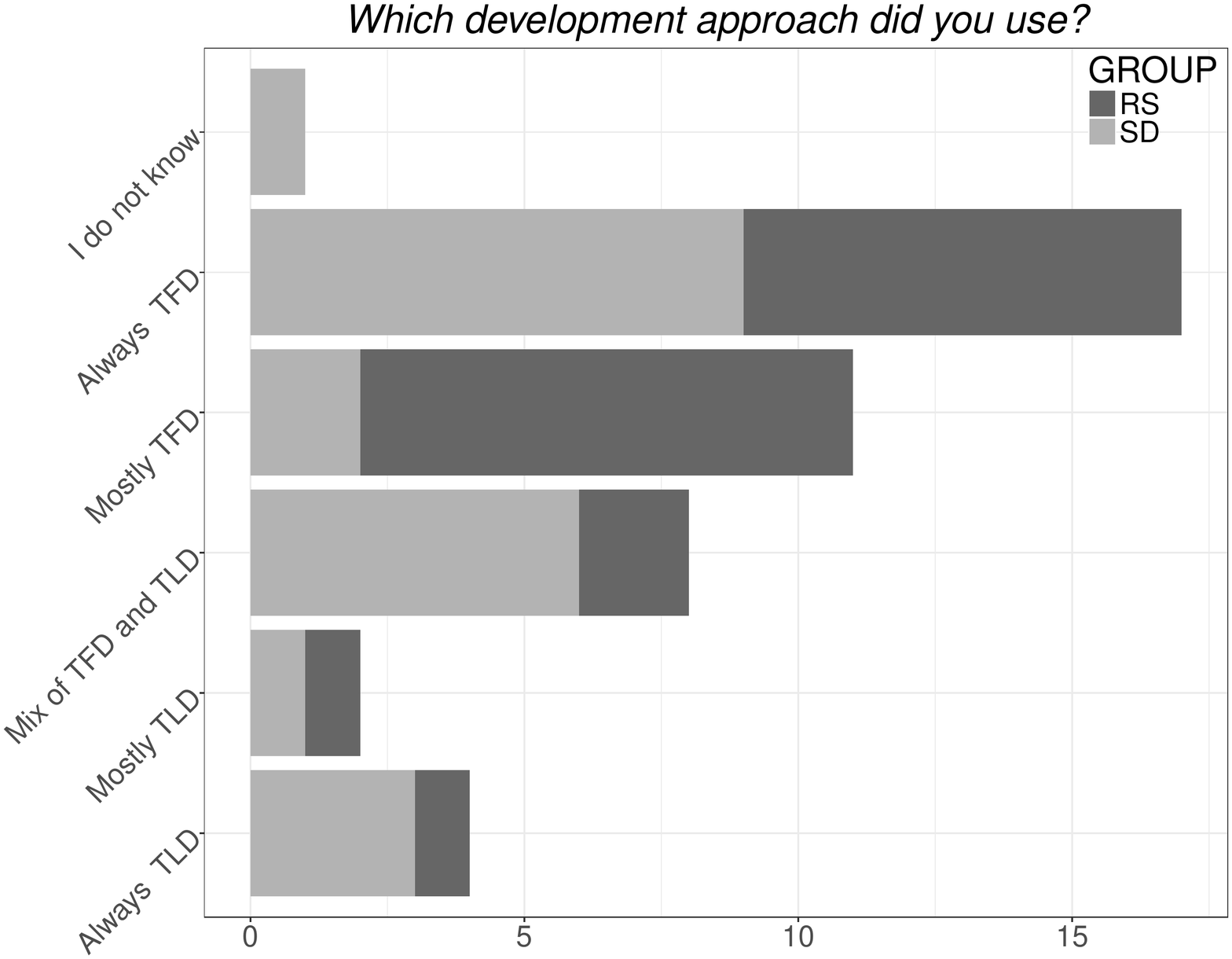}}
}
\caption{Answers to the post-questionnaire questions regarding the use of test-first development (TFD) during the experimental task, divided by group.}
\label{fig:post}
\end{figure*}

After the experimental session, we administered a post-questionnaire to the participants. 
We asked them to rate how difficult it was to apply TFD to the experimental task (Figure~\ref{fig:post}a), and which development approach they used (Figure~\ref{fig:post}b)---TFD or test-last development (TLD).

\newpage

\end{document}